\documentclass[draftcls,11pt,onecolumn,romanappendices]{IEEEtran}
\usepackage{mathpple}
\usepackage{times}
\usepackage{amsmath}  
\usepackage{amssymb} 
\usepackage{mathrsfs}
\usepackage{cite}
\usepackage{comment} 
\usepackage{upref}
\usepackage{amsfonts}
\usepackage{verbatim}
\usepackage[dvipsnames,usenames]{color}
\usepackage{enumerate}
\usepackage{graphicx}
\usepackage{subfigure}
\usepackage{latexsym}
\usepackage{bm}
\usepackage{multirow}
\usepackage{dsfont}
\usepackage{amsthm,xpatch}
\usepackage{mathtools}
\usepackage{bbm}
\usepackage{latexsym}
\usepackage{accents}
\usepackage{eufrak}
\usepackage{psfrag}
\usepackage{color}
\usepackage{times,color}
\usepackage[usenames,dvipsnames,svgnames,table]{xcolor}
\usepackage{pgf}
\usepackage{tikz}
\usetikzlibrary{arrows, automata, positioning, calc}
\usepackage{cancel} 

\newtheorem{lemma}{Lemma}
\newtheorem{theorem}{Theorem}
\newtheorem{remark}{Remark}

\usepackage{algorithm,algpseudocode}

\makeatletter
\newcommand{\removelatexerror}{\let\@latex@error\@gobble}
\makeatletter
\newcommand{\proofpart}[2]{%
	\par
	\addvspace{\medskipamount}%
	\noindent\emph{Part #1: #2}\par\nobreak
	\addvspace{\smallskipamount}%
	\@afterheading
}
\makeatother

\renewcommand{\mathsf}[1]{#1}

\newcommand{\bsf}[1]{\mathbf{#1}}

\newcommand{\brm}[1]{\mathbf{#1}}

\newcommand{\Proof}{{{\it ~\,Proof. }}}
\newcommand{\QED}{\mbox{}\hfill \raisebox{-0.5ex}{$\Box$}}

\theoremstyle{definition}
\newtheorem{example}{Example}

\begin{document}

\pagestyle{empty}

\title{The Role of Coded Side Information in Single-Server Private Information Retrieval}

\author{Anoosheh Heidarzadeh, Fatemeh Kazemi, and Alex Sprintson\thanks{This work was presented in part at the 2018 IEEE Information Theory Workshop, Guangzhou, China, November 2018, and the 2019 IEEE International Symposium on Information Theory, Paris, France, July 2019.}\thanks{The authors are with the Department of Electrical and Computer Engineering, Texas A\&M University, College Station, TX 77843 USA (E-mail: \{anoosheh, fatemeh.kazemi, spalex\}@tamu.edu).}\thanks{This material is based upon work supported by the National Science Foundation under Grants No.~1718658 and 1642983.}}

\maketitle 

\thispagestyle{empty}

\begin{abstract} 
We study the role of coded side information in single-server Private Information Retrieval (PIR). An instance of the single-server PIR problem includes a server that stores a database of $K$ independently and uniformly distributed messages, and a user who wants to retrieve one of these messages from the server. We consider settings in which the user initially has access to a coded side information which includes a linear combination of a subset of $M$ messages in the database. We assume that the identities of the $M$ messages that form the support set of the coded side information as well as the coding coefficients are initially unknown to the server. We consider two different models, depending on whether the support set of the coded side information includes the requested message or not. We also consider the following two privacy requirements: (i) the identities of both the demand and the support set of the coded side information need to be protected, or (ii) only the identity of the demand needs to be protected. For each model and for each of the privacy requirements, we consider the problem of designing a protocol for generating the user's query and the server's answer that enables the user to decode the message they need while satisfying the privacy requirement. We characterize the (scalar-linear) capacity of each setting, defined as the ratio of the number of information bits in a message to the minimum number of information bits downloaded from the server over all (scalar-linear) protocols that satisfy the privacy condition. Our converse proofs rely on new information-theoretic arguments---tailored to the setting of single-server PIR and different from the commonly-used techniques in multi-server PIR settings. We also present novel capacity-achieving scalar-linear protocols for each of the settings being considered. 
\end{abstract}


\section{Introduction}
In the Private Information Retrieval (PIR) problem, there is a user that wishes to privately download a single or multiple messages belonging to a database stored on a single or multiple (non-colluding or colluding) servers. There are two different types of PIR in the literature: \emph{computational} and \emph{information-theoretic}. In the computational PIR (see, e.g.,~\cite{KO1997}), the identity of the requested message(s) must be protected from the server(s), assuming that the server(s) is computationally bounded. Aside from the computational PIR is the information-theoretic PIR, introduced by Chor \emph{et al.}~in~\cite{Chor:PIR1995}, where no such assumption is made on the computational power of the server(s), and the identity of the requested message(s) need to be protected in an information-theoretic sense. The drawback of this strong requirement is that in the single-server case, the user must download the entire database from the server~\cite{Chor:PIR1995}. This has led to an extensive body of work on multi-server information-theoretic PIR (see, e.g.,~\cite{beimel2002breaking,gasarch2004survey, yekhanin2010private,Sun2017,JafarPIR3new,BU2018,shah2014one, CHY2015, tajeddine2016private, extended, BU18, fazeli2015pir, blackburn2016pir, freij2016private}). 

Initiated by the work of Kadhe \emph{et al.} in~\cite{Kadhe2017} and~\cite{KGHERS2017}, the information-theoretic PIR was recently extended to the settings wherein the user has a random subset of messages in the database as side information, and the identities of side information messages are unknown to the server(s)~\cite{Kadhe2017,KGHERS2017,HKGRS:2018,KHSO2019,LG:2018,KKHS32019,Chen2017side,Maddah2018,HKRS2019}. (Some other types of side information, not closely related to our work, were also studied, see, e.g.,\cite{Tandon2017,Wei2017CacheAidedPI,Wei2017FundamentalLO,WU2018}.)  
Three different notions of privacy were considered: (i) \emph{$(\mathsf{W},\mathsf{S})$-privacy}, where both the identities of the requested messages (denoted by the index set $W$) and the identities of the side information messages (denoted by the index set $S$) must be protected~\cite{Kadhe2017,KGHERS2017,Chen2017side,Maddah2018,HKGRS:2018,KHSO2019}; (ii) \emph{joint $\mathsf{W}$-privacy}, where only the identities of the requested messages (and not necessarily the identities of the side information messages) must be protected~\cite{Kadhe2017,KGHERS2017,HKGRS:2018,KHSO2019,KKHS32019,LG:2018}; and (iii) \emph{individual $\mathsf{W}$-privacy}, where the identity of each requested message must be protected individually (but not necessarily jointly)~\cite{HKRS2019}. In single-message PIR, where the user wants to retrieve one message only, the notions of joint and individual $\mathsf{W}$-privacy, referred to as \emph{$\mathsf{W}$-privacy} for brevity, are equivalent. The differences between these two notions of $\mathsf{W}$-privacy in multi-message PIR were studied in~\cite{HKRS2019}.   

\begin{table*}[t]
\caption{Summary of our main results for single-server PIR with coded side information}
\label{tab:1}
\centering
\begin{tabular}{|c|c|c|c|c|}
\hline
Privacy Condition & \multicolumn{2}{c|}{$(\mathsf{W},\mathsf{S})$-Privacy} & \multicolumn{2}{c|}{$\mathsf{W}$-Privacy} \\ \hline
Model & \begin{tabular}[c]{@{}c@{}} $\mathsf{W}\not\in \mathsf{S}$\\ (PIR-PCSI--I)\end{tabular}  & \begin{tabular}[c]{@{}c@{}} $\mathsf{W}\in \mathsf{S}$\\ (PIR-PCSI--II)\end{tabular} & \begin{tabular}[c]{@{}c@{}} $\mathsf{W}\not\in \mathsf{S}$\\ (PIR-CSI--I)\end{tabular} & \begin{tabular}[c]{@{}c@{}} $\mathsf{W}\in \mathsf{S}$\\ (PIR-CSI--II)\end{tabular} \\ \hline
Parameters & $1\leq M\leq K-1$ & $2\leq M\leq K$ & $1\leq M\leq K-1$ & $2\leq M\leq K$ \\ \hline\hline
Capacity & \multirow{3}{*}{\begin{tabular}[c]{@{}c@{}}$(K-M)^{-1}$ \\ (Theorem~\ref{thm:PIRPCSI-I})\end{tabular}} & \begin{tabular}[c]{@{}c@{}}$(K-M+1)^{-1}$ for $M>\frac{K+1}{2}$\\ (Theorem~\ref{thm:PIRPCSI-II}) \\ Open for $M\leq \frac{K+1}{2}$\\ (Lower bound in Theorem~\ref{thm:PIRPCSI-II})\end{tabular} & \multirow{2}{*}{\begin{tabular}[c]{@{}c@{}}$\lceil \frac{K}{M+1}\rceil^{-1}$\\ (Theorem~\ref{thm:PIRCSI-I})\end{tabular}} & \multirow{2}{*}{\begin{tabular}[c]{@{}c@{}}$1$ for $M=2$, $M=K$\\ $\frac{1}{2}$ for $3\leq M\leq K-1$\\ (Theorem~\ref{thm:PIRCSI-II})\end{tabular}} \\ \cline{1-1} \cline{3-3}
\begin{tabular}[c]{@{}c@{}}Scalar-Linear\\ Capacity\end{tabular} &  & \begin{tabular}[c]{@{}c@{}}$(K-M+1)^{-1}$\\ (Theorem~\ref{thm:PIRPCSI-II})\end{tabular} &  &  \\ \hline\hline
\begin{tabular}[c]{@{}c@{}} Achievability \\ Scheme\end{tabular} & \begin{tabular}[c]{@{}c@{}} Specialized \\ GRS Code\end{tabular} & \begin{tabular}[c]{@{}c@{}}Modified Specialized\\ GRS Code\end{tabular} & \begin{tabular}[c]{@{}c@{}} Modified \\ Partition-and-Code\end{tabular} & \begin{tabular}[c]{@{}c@{}} Randomized \\ Selection-and-Code\end{tabular} \\ \hline
\end{tabular}
\end{table*}

In this work, we focus on single-message single-server information-theoretic PIR in the presence of a \emph{coded side information}. We initiated this study in~\cite{HKS2018} and \cite{HKS2019} for the cases in which $\mathsf{W}$-privacy and $(\mathsf{W},\mathsf{S})$-privacy are required, respectively, where $W$ denotes the index of the requested message, and $S$ denotes the index set of the messages in the support set of the coded side information. We have recently extended these works to the multi-server setting in~\cite{KKHS12019} and \cite{KKHS22019}.  
In this problem, there is a single server storing a database of $K$ independently and uniformly distributed messages; and there is a user who is interested in retrieving a single message from the server. The user initially knows a linear coded combination of a subset of $M$ messages in the database, where the identities of the messages in the support set of the user's coded side information as well as their coding coefficients are initially unknown to the server. This setting can be motivated by several practical scenarios. For instance, the user may have obtained a coded side information via overhearing in a wireless network; or on-the-fly recording of a random linear combination of messages being broadcast by an information source; or from a trusted agent, e.g., an entity who makes profit by offering privacy to users,  with limited knowledge about the database; or from the information which is locally stored, e.g., using an erasure code, in the user's cache of limited size. Recently, inspired by~\cite{Kadhe2017}, a group of researchers from Google in~\cite{PPY2018} used the idea of a coded side information in a new single-server PIR scheme, which leverages both the information-theoretic and computational PIR.  

The problem is to design a protocol for generating the user's query and the server's answer which satisfy one of the following two privacy conditions: $(\mathsf{W},\mathsf{S})$-privacy, i.e., the privacy of both the requested message and the messages in the support set of the coded side information must be preserved, or $\mathsf{W}$-privacy, i.e., only the privacy of the requested message needs to be protected. We refer to this problem as \emph{PIR with Private Coded Side Information (PIR-PCSI)} or \emph{PIR with Coded Side Information (PIR-CSI)} when $(\mathsf{W},\mathsf{S})$-privacy or $\mathsf{W}$-privacy is required, respectively. 

Depending on whether the support set of the user's coded side information includes the user's demand or not, we consider two different models for each of the PIR-PCSI and PIR-CSI problems. In the first model, referred to as \emph{Model~I}, the demand does not belong to the support set of the coded side information, whereas in the second model, referred to as \emph{Model~II}, the demand belongs to the support set of the coded side information. We refer to the PIR-PCSI (or PIR-CSI) problem under Model~I and Model~II as \emph{PIR-PCSI--I} (or \emph{PIR-CSI--I}) and \emph{PIR-PCSI--II} (or \emph{PIR-CSI--II}), respectively. 

For each of these settings, we define the capacity as the ratio of the number of information bits in a message to the minimum number of information bits downloaded from the server over all protocols that satisfy the privacy condition. We similarly define the scalar-linear capacity of each setting, except when the minimum is taken over all scalar-linear protocols---the protocols in which the server's answer contains only scalar-linear combinations of the messages in the database (i.e., linear combinations with scalar coding coefficients), that satisfy the privacy condition. In this work, our goal is to characterize the capacity and the scalar-linear capacity of each of the PIR-PCSI and PIR-CSI settings, and design a capacity-achieving protocol for each of these settings.  

The main contributions of this work are as follows. The results are also summarized in Table~\ref{tab:1}.

For the PIR-PCSI--I setting, we prove that the capacity and the scalar-linear capacity are both given by $(K-M)^{-1}$ for any ${1\leq M\leq K-1}$. This is interesting because, as shown in~\cite[Theorem~2]{Kadhe2017}, the capacity of PIR with $M$ randomly chosen messages as side information is equal to $(K-M)^{-1}$ when $(\mathsf{W},\mathsf{S})$-privacy is required. This shows that for achieving $(\mathsf{W},\mathsf{S})$-privacy, even \emph{one} random linear coded combination of a random subset of $M$ messages is as efficient as $M$ randomly chosen messages separately, as side information. 

For the PIR-PCSI--II setting, we prove that the scalar-linear capacity for any value of ${2\leq M\leq K}$ and the capacity for any value of ${\frac{K+1}{2}<M\leq K}$ are given by $(K-M+1)^{-1}$, whereas the capacity for any value of $2\leq M\leq \frac{K+1}{2}$ remains open. This shows that when the user knows only \emph{one} random linear coded combination whose support set consists of the requested message along with $M-1$ other randomly chosen messages, achieving $(\mathsf{W},\mathsf{S})$-privacy is no more costly than that when the user knows $M-1$ randomly chosen (uncoded) messages, different from the requested message.  

For the PIR-CSI--I setting, we prove that the capacity and the scalar-linear capacity are given by $\lceil\frac{K}{M+1}\rceil^{-1}$ for any $0\leq M< K$. Interestingly, this is the same as the capacity of PIR with $M$ randomly chosen messages as side information~\cite[Theorem~1]{Kadhe2017}. For the PIR-CSI--II setting, we prove that the capacity and the scalar-linear capacity are equal to~$1$ for $M=2$ and $M=K$, and are equal to~$\frac{1}{2}$ for any ${3\leq M\leq K-1}$. This result is particularly interesting because, unlike the previous settings, the gap between the capacity and the trivial capacity upper bound $1$ is a constant, regardless of the size of support set of the side information ($M$). 

The converse proofs are based on new information-theoretic arguments. These arguments are tailored to the setting of single-server PIR and are different from the proof techniques being commonly used in the multi-server PIR settings. In particular, the main ingredients in the proofs are a necessary condition for $(\mathsf{W},\mathsf{S})$-privacy and a necessary condition for $\mathsf{W}$-privacy, which reveal the combinatorial nature of the problem of single-server PIR in the presence of (uncoded or coded) side information. In addition, our converse proofs for the PIR-PCSI--I and PIR-CSI--I settings serve as alternative information-theoretic proofs for the results in~\cite{Kadhe2017} which were proven using index coding arguments. 

The achievability proofs are based on novel scalar-linear PIR-PCSI and PIR-CSI protocols. In particular, the proposed PIR-PCSI--I and PIR-PCSI--II protocols, termed the \emph{Specialized GRS Code protocol} and the \emph{Modified Specialized GRS Code protocol}, rely on the Generalized Reed-Solomon (GRS) codes that contain a specific codeword, depending on the index of the requested message as well as the indices of the messages in the support set of the coded side information and their coding coefficients. 

The proposed protocol for the PIR-CSI--I setting, termed the \emph{Modified Partition-and-Code (MPC) protocol}, is inspired by our recently proposed Partition-and-Code with Interference Alignment protocol in~\cite{HS2019} for single-server private computation with uncoded side information. The MPC protocol also generalizes the Partition-and-Code protocol of~\cite{Kadhe2017} for single-server PIR with uncoded side information. It is noteworthy that we originally introduced a different PIR-CSI--I protocol in~\cite{HKS2018}, termed \emph{Randomized Partitioning (RP) protocol}, which is also capacity-achieving. 

For the PIR-CSI--II setting, we propose a protocol, termed the \emph{Randomized Selection-and-Code protocol}, which is based on the idea of randomizing the structure of the user's query and the server's answer (instead of always using a fixed structure for query/answer). We introduced this idea in~\cite{HKS2018} for the first time, and Tian \emph{et al.}, concurrently and independently, used a similar idea in~\cite{TSC2018} for multi-server PIR without side information. 

\section{Problem Setup and Formulation}\label{sec:SN}
\subsection{Basic Notation}
Throughout this paper, we denote random variables and their realizations by bold-face and regular letters, respectively. The functions $\mathbb{P}(\cdot)$, $\mathbb{P}(\cdot|\cdot)$, $H(\cdot)$, $H(\cdot|\cdot)$, and $I(\cdot;\cdot |\cdot)$ denote probability, conditional probability, (Shannon) entropy, conditional entropy, and conditional mutual information, respectively.

Let $\mathbb{F}_q$ be a finite field for a prime power $q$, and let $\mathbb{F}^{\times}_q\triangleq \mathbb{F}_q\setminus \{0\}$ be the multiplicative group of $\mathbb{F}_q$. Let $\mathbb{F}_{q^l}$ be an extension field of $\mathbb{F}_q$ for an integer $l\geq 1$, and let $L\triangleq l\log_2 q$. The parameters $q$ and $l$ are referred to as the \emph{base-field size} and the \emph{field-extension degree}, respectively.

Let $K\geq 1$ and $1\leq M\leq K$ be two integers. Let $\mathcal{K} \triangleq \{1,\dots,K\}$. We denote by $\mathcal{S}$ the set of all $M$-subsets (i.e., all subsets of size $M$) of $\mathcal{K}$, and denote by $\mathcal{C}$ the set of all ordered multisets of size $M$ (i.e., all length-$M$ sequences) with elements from $\mathbb{F}_q^{\times}$. Note that $|\mathcal{S}| = \binom{K}{M}$ and $|\mathcal{C}| = (q-1)^{M}$. 

\subsection{Setup and Assumptions}
There is a server that stores a set of $K$ messages ${X}_1,\dots,{X}_K$, denoted by $X_{\mathcal{K}}\triangleq \{X_1,\dots,X_K\}$, where $\brm{X}_i$'s are independently and uniformly distributed over $\mathbb{F}_{q^l}$, i.e., ${H(\brm{X}_i) = L}$ for $i\in \mathcal{K}$ and $H(\brm{X}_{\mathcal{K}}) = KL$, where $\brm{X}_{\mathcal{K}}\triangleq \{\brm{X}_1,\dots,\brm{X}_K\}$. There is a user who wants to retrieve a message ${X}_{\mathsf{W}}$ for some $\mathsf{W}\in \mathcal{K}$ from the server, and knows a linear combination ${Y}^{[\mathsf{S},\mathsf{C}]}\triangleq \sum_{i\in \mathsf{S}} {c}_i {X}_i$ on the messages $X_{\mathsf{S}}\triangleq \{X_i: i\in S\}$, for some $\mathsf{S} \triangleq \{i_1,\dots,i_M\}\in \mathcal{S}$ and ${\mathsf{C} \triangleq \{{c}_{i_1},\dots,{c}_{i_M}\} \in \mathcal{C}}$. We refer to ${X}_{\mathsf{W}}$ as the \emph{demand}, $\mathsf{W}$ as the \emph{demand index}, $X_{\mathsf{S}}$ as the \emph{side information support set},  $\mathsf{S}$ as the \emph{side information support index set}, $M$ as the \emph{side information support size}, and ${Y}^{[\mathsf{S},\mathsf{C}]}$ as the \emph{(coded) side information}. 

We assume that $\bsf{S}$ and $\bsf{C}$ are uniformly distributed over $\mathcal{S}$ and $\mathcal{C}$, respectively. Also, we consider two different models for the conditional distribution of $\bsf{W}$ given $\bsf{S}=\mathsf{S}$: 

\subsubsection*{Model~I} $\bsf{W}$ is uniformly distributed over $\mathcal{K}\setminus\mathsf{S}$, 
\begin{equation*}
\mathbb{P}(\bsf{W}=\mathsf{W}|\bsf{S}=\mathsf{S}) = 
\begin{cases}
\frac{1}{K-M}, & \mathsf{W}\in \mathcal{K}\setminus\mathsf{S},\\	
0, & \text{otherwise};
\end{cases}
\end{equation*} 

\subsubsection*{Model~II} $\bsf{W}$ is uniformly distributed over $\mathsf{S}$, 
\begin{equation*}
\mathbb{P}(\bsf{W}=\mathsf{W}|\bsf{S}=\mathsf{S}) = 
\begin{cases}
\frac{1}{M}, & \mathsf{W}\in \mathsf{S},\\	
0, & \text{otherwise}.
\end{cases}
\end{equation*} For both Models~I and~II, $\bsf{W}$ is distributed uniformly over $\mathcal{K}$. 

Let $\mathds{1}_{\{\mathbf{W}\in \mathbf{S}\}}$ be an indicator random variable such that that $\mathds{1}_{\{\mathbf{W}\in \mathbf{S}\}}=0$ if $\mathbf{W}\not\in \mathbf{S}$, and $\mathds{1}_{\{\mathbf{W}\in \mathbf{S}\}}=1$ otherwise. Note that $\mathds{1}_{\{\mathbf{W}\in \mathbf{S}\}} = 0$ in Model I, and $\mathds{1}_{\{\mathbf{W}\in \mathbf{S}\}} = 1$ in Model II. 

We assume that the server knows the underlying model (i.e., whether $\bsf{W}\not\in \bsf{S}$ or $\bsf{W}\in \bsf{S}$), the side information support size $M$, the distributions of $\bsf{S}$ and $\bsf{C}$, and the conditional distribution of $\bsf{W}$ given $\bsf{S}$, in advance; whereas the realizations $\mathsf{W},\mathsf{S},\mathsf{C}$ are unknown to the server in advance. 

\subsection{Privacy and Recoverability Conditions}
For any given $\mathsf{W}, \mathsf{S}, \mathsf{C}$, in order to retrieve ${X}_{\mathsf{W}}$, the user sends to the server a query ${Q}^{[\mathsf{W},\mathsf{S},\mathsf{C}]}$, which is a (potentially stochastic) function of $\mathsf{W},\mathsf{S},\mathsf{C}$.\footnote{In general, the query may also depend on the content of the side information---notwithstanding, in this work we focus on queries that are ``universal'' in the sense that any such query achieves privacy for all realizations of the messages.} For simplifying the notation, we denote $\brm{Q}^{[\bsf{W},\bsf{S},\bsf{C}]}$ by $\brm{Q}$. 

The query must satisfy one of the following two privacy conditions: 
\begin{itemize}
\item[(i)] both the user's demand index and side information support index set must be protected from the server; 
\item[(ii)] 	only the user's demand index (and not necessarily the side information support index set) must be protected from the server.
\end{itemize} The condition~(i) is referred to as the \emph{$(\mathsf{W},\mathsf{S})$-privacy condition}, and the condition~(ii) is referred to as the \emph{$\mathsf{W}$-privacy condition}. (Note that $(\mathsf{W},\mathsf{S})$-privacy is a stronger condition than $\mathsf{W}$-privacy.) The $(\mathsf{W},\mathsf{S})$-privacy condition implies that $(\bsf{W},\bsf{S})$ and $\mathbf{Q}$ must be conditionally independent given $\mathds{1}_{\{\bsf{W}\in\bsf{S}\}}$, \[I(\bsf{W},\bsf{S};\brm{Q}|\mathds{1}_{\{\bsf{W}\in\bsf{S}\}}) = 0.\] The $\mathsf{W}$-privacy condition implies that $\bsf{W}$ and $\mathbf{Q}$ must be conditionally independent given $\mathds{1}_{\{\bsf{W}\in\bsf{S}\}}$, \[I(\bsf{W};\brm{Q}|\mathds{1}_{\{\bsf{W}\in\bsf{S}\}}) = 0.\] Equivalently, for a given $\theta\in \{0,1\}$, when $(\mathsf{W},\mathsf{S})$-privacy is required, it must hold that 
\begin{align*}& \mathbb{P}(\bsf{W}= \mathsf{W}^{*},\bsf{S}=\mathsf{S}^{*}| \brm{Q}= {Q}^{[\mathsf{W},\mathsf{S},\mathsf{C}]},\mathds{1}_{\{\bsf{W}\in \bsf{S}\}}=\theta) \\ & = \mathbb{P}(\bsf{W}= \mathsf{W}^{*},\bsf{S}=\mathsf{S}^{*}|\mathds{1}_{\{\bsf{W}\in \bsf{S}\}}=\theta)\end{align*} for all $\mathsf{W}^{*}\in \mathcal{K}$ and $\mathsf{S}^{*}\in \mathcal{S}$, and when $\mathsf{W}$-privacy is required, it must hold that
\begin{align*}&\mathbb{P}({\bsf{W}=\mathsf{W}^{*}}|{\brm{Q} = {Q}^{[\mathsf{W},\mathsf{S},\mathsf{C}]}},{\mathds{1}_{\{\bsf{W}\in \bsf{S}\}}=\theta}) \\ & = {\mathbb{P}(\bsf{W}=\mathsf{W}^{*}|{\mathds{1}_{\{\bsf{W}\in \bsf{S}\}}=\theta})}
\end{align*} for all $\mathsf{W}^{*}\in \mathcal{K}$.\footnote{The mutual information based definitions of the $(W,S)$-privacy and $W$-privacy conditions will be used in the converse proofs, whereas their probability based counterparts will be used in the achievability proofs.}

Upon receiving ${Q}^{[\mathsf{W},\mathsf{S},\mathsf{C}]}$, the server sends to the user an answer ${A}^{[\mathsf{W},\mathsf{S},\mathsf{C}]}$, which is a (deterministic) function of the query ${Q}^{[\mathsf{W},\mathsf{S},\mathsf{C}]}$, the indicator variable $\mathds{1}_{\{\mathsf{W}\in \mathsf{S}\}}$, and the messages in $X_{\mathcal{K}}$. For simplifying the notation, we denote $\brm{A}^{[\bsf{W},\bsf{S},\bsf{C}]}$ by $\brm{A}$. Note that
$(\bsf{W},\bsf{S},\bsf{C}) \leftrightarrow (\brm{Q},\mathds{1}_{\{\bsf{W}\in \bsf{S}\}},\brm{X}_{\mathcal{K}}) \leftrightarrow \brm{A}$ forms a Markov chain, and 
$H(\brm{A}| \brm{Q},\mathds{1}_{\{\bsf{W}\in \bsf{S}\}}, \brm{X}_{\mathcal{K}},\bsf{W},\bsf{S},\bsf{C}) = 0$. 

The answer ${A}^{[\mathsf{W},\mathsf{S},\mathsf{C}]}$ along with ${Q}^{[\mathsf{W},\mathsf{S},\mathsf{C}]}, \mathds{1}_{\{\mathsf{W}\in \mathsf{S}\}}, Y^{[\mathsf{S},\mathsf{C}]}$, and $\mathsf{W}, \mathsf{S}, \mathsf{C}$ must enable the user to retrieve the demand ${X}_{\mathsf{W}}$. That is, it must hold that \[H(\brm{X}_{\bsf{W}}| \brm{A}, \brm{Q}, \mathds{1}_{\{\bsf{W}\in \bsf{S}\}},\brm{Y}^{[\bsf{S},\bsf{C}]}, \bsf{W},\bsf{S},\bsf{C})=0.\] We refer to this condition as the \emph{recoverability condition}. 

\subsection{PIR-PCSI and PIR-CSI Problems}
For each type of privacy and for each model, the problem is to design a protocol for generating a query ${Q}^{[\mathsf{W},\mathsf{S},\mathsf{C}]}$ (and the corresponding answer ${A}^{[\mathsf{W},\mathsf{S},\mathsf{C}]}$, given ${Q}^{[\mathsf{W},\mathsf{S},\mathsf{C}]}$, $\mathds{1}_{\{\mathsf{W}\in \mathsf{S}\}}$, and ${X}_{\mathcal{K}}$) for any given $\mathsf{W},\mathsf{S},\mathsf{C}$, such that both the privacy and recoverability conditions are satisfied. Note that the protocol is assumed to be known at the server. When $(\mathsf{W},\mathsf{S})$-privacy is required, we refer to this problem as \emph{Private Information Retrieval (PIR) with Private Coded Side Information} \emph{(PIR-PCSI)}, and when $\mathsf{W}$-privacy is required we refer to this problem as \emph{PIR with Coded Side Information} \emph{(PIR-CSI)}. 

The PIR-PCSI problem under Model~I (or Model~II) is referred to as the \emph{PIR-PCSI--I} (or \emph{PIR-PCSI--II}) setting; and the PIR-CSI problem under Model~I (or Model~II) is referred to as the \emph{PIR-CSI--I} (or \emph{PIR-CSI--II}) setting. A protocol for generating query/answer for the PIR-PCSI--I (or PIR-PCSI--II) setting is referred to as a \emph{PIR-PCSI--I} (or \emph{PIR-PCSI--II}) \emph{protocol}. A \emph{PIR-CSI--I} (or \emph{PIR-CSI--II}) \emph{protocol} is defined similarly.   

\subsection{Capacity and Scalar-Linear Capacity}
The \emph{rate of a PIR-PCSI--I (or PIR-PCSI--II) protocol} is defined as the ratio of the entropy of a message, i.e., $L$, to the conditional entropy of $\brm{A}^{[\bsf{W},\bsf{S},\bsf{C}]}$ given that $\mathds{1}_{\{\bsf{W}\in \bsf{S}\}}=0$ (or $\mathds{1}_{\{\bsf{W}\in \bsf{S}\}}=1$). The \emph{rate of a PIR-CSI--I (or PIR-CSI--II) protocol} is defined similarly. 

The \emph{capacity of PIR-PCSI--I (or PIR-PCSI--II) setting} is defined as the supremum of rates over all PIR-PCSI--I (or PIR-PCSI--II) protocols and over all base-field sizes $q$ and all field-extension degrees $l$; and the \emph{capacity of PIR-CSI--I (or PIR-CSI--II) setting} is defined similarly. The \emph{scalar-linear capacity of PIR-PCSI--I (or PIR-PCSI--II) setting} is defined as the supremum of rates over all scalar-linear PIR-PCSI--I (or PIR-PCSI--II) protocols (i.e., the protocols in which the answer of the server consists only of the scalar-linear combinations of the messages in $X_{\mathcal{K}}$) and over all $q$ and $l$. The \emph{scalar-linear capacity of PIR-CSI--I (or PIR-CSI--II) setting} is defined similarly.\footnote{Although our definitions of capacity and scalar-linear capacity are independent of the base-field size $q$ and the field-extension degree $l$, these quantities may depend on $q$ and $l$ in general. In this work, we show that the capacity and the scalar-linear capacity of the PIR-PCSI settings are achievable so long as $q\geq K$ and $l\geq 1$; and depending on the parameters $K,M$ and the model (I or~II), the capacity and the scalar-linear capacity of the PIR-CSI settings are achievable so long as $q\geq 2$ or $q\geq 3$ and $l\geq 1$.}  

\subsection{Problem Statement}
In this work, our goal is to derive upper bounds on the capacity and the scalar-linear capacity of the PIR-PCSI--I, PIR-PCSI--II, PIR-CSI--I, and PIR-CSI--II settings, and to design protocols that achieve the corresponding upper-bounds. 

\section{Main Results}
We present our main results in this section. The results for the PIR-PCSI--I and PIR-PCSI--II settings are summarized in Section~\ref{subsec:PCSIMainResults}, and the results for the PIR-CSI--I and PIR-CSI--II settings are summarized in Section~\ref{subsec:CSIMainResults} 

The following two lemmas give a necessary condition for $(\mathsf{W},\mathsf{S})$-privacy and $\mathsf{W}$-privacy, respectively. These simple but powerful lemmas are the key components in the converse proofs of our main results. 

\begin{lemma}\label{prop:1}
For $(\mathsf{W},\mathsf{S})$-privacy, for a given ${\theta\in \{0,1\}}$, for any $\mathsf{W}^{*}\in\mathcal{K}$ and $\mathsf{S}^{*}\in\mathcal{S}$ with ${\mathds{1}_{\{\mathsf{W}^{*}\in \mathsf{S}^{*}\}}=\theta}$, there must exist ${\mathsf{C}^{*}\in\mathcal{C}}$ such that \[H(\brm{X}_{\mathsf{W}^{*}}| \brm{A}, \brm{Q}, \mathds{1}_{\{\bsf{W}\in \bsf{S}\}}=\theta, \brm{Y}^{[\mathsf{S}^{*},\mathsf{C}^{*}]}) = 0.\] 
\end{lemma}

\Proof
The proof is by the way of contradiction. For a given $\theta\in \{0,1\}$, consider an arbitrary $\mathsf{W}^{*}\in \mathcal{K}$ and an arbitrary $\mathsf{S}^{*}\in \mathcal{S}$ such that $\mathds{1}_{\{\mathsf{W}^{*}\in \mathsf{S}^{*}\}}=\theta$. Suppose that there does not exist any $\mathsf{C}^{*}\in \mathcal{C}$ such that ${H(\brm{X}_{\mathsf{W}^{*}}| \brm{A}, \brm{Q}, \mathds{1}_{\{\bsf{W}\in \bsf{S}\}}=\theta, \brm{Y}^{[\mathsf{S}^{*},\mathsf{C}^{*}]}) = 0}$. If $\mathsf{W}^{*}$ and $\mathsf{S}^{*}$ are respectively the user's demand index and side information support index set, no matter what the user's side information $Y^{[\mathsf{S}^{*},\cdot]}$ is, the user cannot recover $X_{\mathsf{W}^{*}}$ given the answer, query, and the side information $Y^{[\mathsf{S}^{*},\cdot]}$. This violates the recoverability condition. Thus, $\mathsf{W}^{*}$ and $\mathsf{S}^{*}$ cannot be the user's demand index and side information support index set, respectively. This obviously violates the $(\mathsf{W},\mathsf{S})$-privacy condition, because given the query, every $\mathsf{W}^{*}\in \mathcal{K}$ and every $\mathsf{S}^{*}\in \mathcal{S}$ such that $\mathds{1}_{\{\mathsf{W}^{*}\in \mathsf{S}^{*}\}}=\theta$ must be equally likely to be the user's demand index and side information support index set, respectively.\QED

\begin{lemma}\label{prop:2}
For $\mathsf{W}$-privacy, for a given ${\theta\in \{0,1\}}$, for any ${\mathsf{W}^{*}\in\mathcal{K}}$, there must exist $\mathsf{S}^{*}\in\mathcal{S}$ with ${\mathds{1}_{\{\mathsf{W}^{*}\in \mathsf{S}^{*}\}}=\theta}$ and ${\mathsf{C}^{*}\in\mathcal{C}}$ such that \[H(\brm{X}_{\mathsf{W}^{*}}| \brm{A}, \brm{Q}, \mathds{1}_{\{\bsf{W}\in \bsf{S}\}}=\theta, \brm{Y}^{[\mathsf{S}^{*},\mathsf{C}^{*}]}) = 0.\] 
\end{lemma}

\Proof
The proof is similar to the proof of Lemma~\ref{prop:1}---except that the $\mathsf{W}$-privacy condition is used instead of the $(\mathsf{W},\mathsf{S})$-privacy condition. The proof is omitted for brevity.\QED

\subsection{PIR-PCSI}\label{subsec:PCSIMainResults}
In this section, we present our main results for the PIR-PCSI--I and PIR-PCSI--II settings. The capacity and the scalar-linear capacity of the PIR-PCSI--I setting (for all ${1\leq M\leq K-1}$) are characterized in Theorem~\ref{thm:PIRPCSI-I}, and the capacity (for all $\frac{K+1}{2}<M\leq K$) and 
the scalar-linear capacity (for all $2\leq M\leq K$) of the PIR-PCSI--II setting are characterized in Theorem~\ref{thm:PIRPCSI-II}. For any $2\leq M\leq \frac{K+1}{2}$, the capacity of the PIR-PCSI--II setting, which we conjecture to be the same as the scalar-linear capacity, remains open. The proofs are given in Sections~\ref{sec:PIRPCSI-I} and~\ref{sec:PIRPCSI-II}. 

\begin{theorem}\label{thm:PIRPCSI-I}
For the PIR-PCSI--I setting with $K$ messages and side information support size $M$, the capacity and the scalar-linear capacity are given by $(K-M)^{-1}$ for all $1\leq M\leq K-1$.
\end{theorem}

The converse follows directly from the result of~\cite[Theorem~2]{Kadhe2017}, which was proven using an index coding argument, for single-server single-message PIR with (uncoded) side information when $(\mathsf{W},\mathsf{S})$-privacy is required. In this work, we provide an alternative proof of converse by upper bounding the rate of any PIR-PCSI--I protocol using the information-theoretic arguments (see Section~\ref{subsec:ConvThm1}). The key component of the proof is the necessary condition for $(\mathsf{W},\mathsf{S})$-privacy, stated in Lemma~\ref{prop:1}. 

The achievability proof relies on a new scalar-linear PIR-PCSI--I protocol, termed the \emph{Specialized GRS Code protocol}, which achieves the rate $(K-M)^{-1}$ (see Section~\ref{subsec:AchThm1}). This protocol is based on the Generalized Reed-Solomon (GRS) codes that contain a specific codeword depending on $\mathsf{W}, \mathsf{S},\mathsf{C}$.  

\begin{remark}\label{rem:1}
\emph{As shown in~\cite{Kadhe2017}, when there is a single server storing $K$ independent and identically distributed messages, and there is a user that knows $M$ randomly chosen (uncoded) messages as their side information and demands a single message not in their side information, in order to guarantee $(\mathsf{W},\mathsf{S})$-privacy, the minimum download cost is $(K-M)L$, where $L$ is the entropy of a message. Surprisingly, this result matches the result of Theorem~\ref{thm:PIRPCSI-I}. This shows that, when compared to having $M$ random messages separately as side information, for achieving $(\mathsf{W},\mathsf{S})$-privacy there will be no additional loss in capacity even if only \emph{one} random linear coded combination of $M$ random messages is known by the user.}
\end{remark}

\begin{theorem}\label{thm:PIRPCSI-II}
For the PIR-PCSI--II setting with $K$ messages and side information support size $M$, the capacity is given by ${(K-M+1)^{-1}}$ for all $\frac{K+1}{2}<M\leq K$, and it is lower bounded by ${(K-M+1)^{-1}}$ for all $2\leq M\leq \frac{K+1}{2}$. Moreover, the scalar-linear capacity is given by $(K-M+1)^{-1}$ for all ${2\leq M\leq K}$. 
\end{theorem}

The converse proof for the scalar-linear case is based on a mix of algebraic and information-theoretic arguments (see Section~\ref{subsec:ConvThm2}), and the proof of converse for the general case relies on different information-theoretic arguments. The main ingredient of the proofs is the result of Lemma~\ref{prop:1}. 

The proof of achievability is based on a novel scalar-linear protocol, referred to as the \emph{Modified Specialized GRS Code protocol}---a modified version of the Specialized GRS Code protocol, which achieves the rate ${(K-M+1)^{-1}}$ (see Section~\ref{subsec:AchThm2}). 

\begin{remark}\label{rem:2}
\emph{Interestingly, comparing the results of~\cite[Theorem~2]{Kadhe2017} and Theorem~\ref{thm:PIRPCSI-II}, one can see that when the side information is composed of $M-1$ randomly chosen messages (different from the requested message), $(\mathsf{W},\mathsf{S})$-privacy cannot be achieved more efficiently than the case in which the side information is only \emph{one} random linear coded combination of $M$ randomly chosen messages including the demand.}
\end{remark}

\subsection{PIR-CSI}\label{subsec:CSIMainResults}
In this section, we present our main results for the PIR-CSI--I and PIR-CSI--II settings. The capacity and the scalar-linear capacity of the PIR-CSI--I setting (for all ${1\leq M\leq K-1}$) and the capacity and the scalar-linear capacity of the PIR-CSI--II setting (for all ${2\leq M\leq K}$) are characterized in Theorems~\ref{thm:PIRCSI-I} and~\ref{thm:PIRCSI-II}, respectively. The proofs are given in Sections~\ref{sec:PIRCSI-I} and~\ref{sec:PIRCSI-II}.

\begin{theorem}\label{thm:PIRCSI-I}
For the PIR-CSI--I setting with $K$ messages and side information support size $M$, the capacity and the scalar-linear capacity are given by ${\lceil \frac{K}{M+1} \rceil}^{-1}$ for all $1\leq M\leq K-1$. 
\end{theorem}

The proof consists of two parts. In the first part, using information-theoretic arguments, we give an upper bound on the rate of any PIR-CSI--I protocol (see Section~\ref{subsec:ConvThm3}). The proofs rely primarily on the necessary condition for $\mathsf{W}$-privacy, stated in Lemma~\ref{prop:2}. In the second part, we construct a new scalar-linear PIR-CSI--I protocol, termed the \emph{Modified Partition-and-Code (MPC) protocol}, which achieves this rate upper-bound (see Section~\ref{subsec:AchThm3}). The proposed protocol is inspired by our recently proposed Partition-and-Code with Interference Alignment protocol in~\cite{HS2019} for single-server private computation with uncoded side information.  

\begin{remark}\label{rem:3}
\emph{Interestingly, the capacity of PIR with (uncoded) side information~\cite{Kadhe2017} is also equal to $\lceil\frac{K}{M+1}\rceil^{-1}$ where $M$ is the number of (uncoded) messages known to the user in advance as side information. This shows that there will be no loss in capacity, when compared to the case that the user knows $M$ randomly chosen messages separately, even if the user knows only \emph{one} random linear coded combination of $M$ randomly chosen messages.} 
\end{remark}

\begin{remark}\label{rem:4}
\emph{When $(\mathsf{W},\mathsf{S})$-privacy is required, the result of Theorem~\ref{thm:PIRPCSI-I} shows that the capacity of single-server PIR with a coded side information with support size $M$ that does not include the demand is equal to $(K-M)^{-1}$. Note that $\lceil\frac{K}{M+1}\rceil< K-M$ for all ${1\leq M\leq K-2}$. This implies that the capacity of the PIR-CSI--I setting is strictly greater than that of the PIR-PCSI--I setting for any ${1\leq M\leq K-2}$. This is expected because $\mathsf{W}$-privacy is a weaker notion of privacy when compared to $(\mathsf{W},\mathsf{S})$-privacy. However, for the extremal case of $M=K-1$, as can be seen $(\mathsf{W},\mathsf{S})$-privacy comes at no extra cost compared to $\mathsf{W}$-privacy.}
\end{remark}

\begin{theorem}\label{thm:PIRCSI-II}
For the PIR-CSI--II setting with $K$ messages and side information support size $M$, the capacity and the scalar-linear capacity are equal to $1$ for $M=2,K$, and $1/2$ for all $3\leq M\leq K-1$. 
\end{theorem}

For each range of values of $M$, the proof consists of two parts. In the first part, we use information-theoretic arguments---based on the result of Lemma~\ref{prop:2}, so as to upper bound the rate of any PIR-CSI--II protocol (see Section~\ref{subsec:ConvThm4}). In the second part, we construct novel scalar-linear {PIR-CSI--II} protocols, collectively termed the \emph{Randomized Selection-and-Code (RSC) protocols}, for different ranges of values of $M$. The proposed protocols rely on probabilistic techniques, and achieve the corresponding rate upper-bounds (see Section~\ref{subsec:AchThm4}). 

\begin{remark}\label{rem:5}
\emph{Interestingly, Theorem~\ref{thm:PIRCSI-II} shows that when $\mathsf{W}$-privacy is required, no matter what the size of support set of the side information is, the user can privately retrieve any message belonging to the support set of their coded side information, with a download cost at most twice the cost of downloading the message directly---which obviously does not preserve the privacy of the requested message.} 
\end{remark}

\begin{remark}\label{rem:6}
\emph{As shown in Theorem~\ref{thm:PIRPCSI-II}, when $(\mathsf{W},\mathsf{S})$-privacy is required, the (scalar-linear) capacity of single-server PIR with a coded side information whose support set includes the requested message is equal to $(K-M+1)^{-1}$, where $M$ is the side information support size. The result of Theorem~\ref{thm:PIRCSI-II} matches this result for the cases of $M=K$ and $M=K-1$, and hence, $(\mathsf{W},\mathsf{S})$-privacy and $\mathsf{W}$-privacy are attainable at the same cost in these cases; whereas for the other cases of $M$, achieving $(\mathsf{W},\mathsf{S})$-privacy is much more costly than achieving $\mathsf{W}$-privacy.}
\end{remark}

\section{Proof of Theorem~\ref{thm:PIRPCSI-I}}\label{sec:PIRPCSI-I}

\subsection{Converse}\label{subsec:ConvThm1}
As shown in~\cite{Kadhe2017} using an index-coding argument, when $(\mathsf{W},\mathsf{S})$-privacy is required, the capacity of PIR with $M$ uncoded messages as side information is given by ${(K-M)^{-1}}$. Obviously, the capacity of the PIR-PCSI--I setting is upper bounded by this quantity. This proves the converse for Theorem~\ref{thm:PIRPCSI-I}. In this section, we present an alternative information-theoretic proof for the general case, which also proves the converse for the scalar-linear case. 

\begin{lemma}\label{lem:Conv1}
For any $1\leq M\leq K-1$, the (scalar-linear) capacity of the PIR-PCSI--I setting is upper bounded by ${(K-M)^{-1}}$.
\end{lemma}

\Proof
In the following, all entropies are conditional on the event $\mathds{1}_{\{\bsf{W}\in \bsf{S}\}}=0$, and we remove this event from the conditions everywhere, for the ease of notation. We need to show that ${H(\brm{A})\geq (K-M)L}$. 

Take arbitrary $\mathsf{W},\mathsf{S},\mathsf{C}$ (and $\brm{Y}\triangleq \brm{Y}^{[\mathsf{S},\mathsf{C}]}$) such that $\mathsf{W}\not\in \mathsf{S}$. Then, we have  
\begin{align}
H(\brm{A}) &\geq H(\brm{A}|\brm{Q},\brm{Y}) \label{eq:PCSIlineI1}\\ 
& = H(\brm{A}|\brm{Q},\brm{Y})+H(\brm{X}_{\mathsf{W}}|\brm{A},\brm{Q},\brm{Y}) \label{eq:PCSIlineI2}\\
&=H(\brm{A},\brm{X}_{\mathsf{W}}|\brm{Q},\brm{Y})\label{eq:PCSIlineI3}\\
&=H(\brm{X}_{\mathsf{W}}|\brm{Q},\brm{Y}) + H(\brm{A}|\brm{Q},\brm{Y},\brm{X}_{\mathsf{W}}) \label{eq:PCSIlineI4}\\
&=H(\brm{X}_{\mathsf{W}})+H(\brm{A}|\brm{Q},\brm{Y},\brm{X}_{\mathsf{W}})\label{eq:PCSIlineI5}
\end{align} where~\eqref{eq:PCSIlineI1} follows since conditioning does not increase the entropy;~\eqref{eq:PCSIlineI2} holds because $H(\brm{X}_{\mathsf{W}}|\brm{A},\brm{Q},\brm{Y})=0$ (by the recoverability condition);~\eqref{eq:PCSIlineI3} and~\eqref{eq:PCSIlineI4} follow from the chain rule of entropy; and~\eqref{eq:PCSIlineI5} follows from $H(\brm{X}_{\mathsf{W}}|\brm{Q},\brm{Y}) = H(\brm{X}_{\mathsf{W}})$ since $\brm{X}_{\mathsf{W}}$ is independent of $(\brm{Q},\brm{Y})$ (noting that $\mathsf{W}\not\in \mathsf{S}$).

If $\mathsf{W}\cup \mathsf{S} = \mathcal{K}$ (i.e., $M=K-1$), then $H(\brm{A})\geq H(\brm{X}_{\mathsf{W}})=L$ (by using the first term in~\eqref{eq:PCSIlineI5}), as was to be shown. If $\mathsf{W}\cup \mathsf{S} \neq \mathcal{K}$, we proceed by lower bounding the second term in~\eqref{eq:PCSIlineI5}, $H(\brm{A}|\brm{Q},\brm{Y},\brm{X}_{\mathsf{W}})$. By Lemma~\ref{prop:1}, for each ${i\in \mathcal{K}\setminus (\mathsf{W}\cup \mathsf{S})}$, there exists $\mathsf{C}_{i}\in \mathcal{C}$ (and $\brm{Y}_i\triangleq \brm{Y}^{[\mathsf{S},\mathsf{C}_i]}$) such that $H(\brm{X}_{i}|\brm{A},\brm{Q},\brm{Y}_{i})=0$. Let $I$ be a maximal subset of ${\mathcal{K}\setminus (\mathsf{W}\cup \mathsf{S})}$ such that $\brm{Y}$ and $\brm{Y}_{I}\triangleq \{\brm{Y}_{i}\}_{i\in I}$ are linearly independent. (Note that ${|I|\leq |\mathsf{S}|-1=M-1}$.) Let $\brm{X}_{I}\triangleq \{\brm{X}_i\}_{i\in I}$. Then, we have
\begin{align}
H(\mathbf{A}|\mathbf{Q},\mathbf{Y},\mathbf{X}_{\mathsf{W}}) & \geq H(\brm{A}|\brm{Q},\brm{Y},\brm{X}_{\mathsf{W}},\brm{Y}_I) \nonumber\\
&\geq H(\mathbf{A}|\mathbf{Q},\mathbf{Y},\mathbf{X}_{\mathsf{W}},\mathbf{Y}_I) \nonumber \\ 
&\quad + H(\mathbf{X}_I|\mathbf{A},\mathbf{Q},\mathbf{Y},\mathbf{X}_{\mathsf{W}},\mathbf{Y}_I)\label{eq:PCSIlineI6}\\ 
&= H(\mathbf{A},\mathbf{X}_I|\mathbf{Q},\mathbf{Y},\mathbf{X}_{\mathsf{W}},\mathbf{Y}_I)\nonumber\\  
& = H(\mathbf{X}_I|\mathbf{Q},\mathbf{Y},\mathbf{X}_{\mathsf{W}},\mathbf{Y}_I)\nonumber\\
& \quad +  H(\mathbf{A}|\mathbf{Q},\mathbf{Y},\mathbf{X}_{\mathsf{W}},\mathbf{Y}_I,\mathbf{X}_I)\nonumber\\
& = H(\mathbf{X}_I) \nonumber \\ 
& \quad +H(\mathbf{A}|\mathbf{Q},\mathbf{Y},\mathbf{X}_{\mathsf{W}},\mathbf{Y}_I,\mathbf{X}_I)\label{eq:PCSIlineI7}	
\end{align} where~\eqref{eq:PCSIlineI6} holds because $H(\mathbf{X}_{i}|\mathbf{A},\mathbf{Q},\mathbf{Y}_{i})=0$ for all $i\in I$ (by assumption); and~\eqref{eq:PCSIlineI7} holds since $\mathbf{X}_I$ is independent of $(\mathbf{Q},\mathbf{Y},\mathbf{X}_{\mathsf{W}},\mathbf{Y}_I)$ by construction (noting that $I$ and ${\mathsf{W}\cup \mathsf{S}}$ are disjoint). The first term in~\eqref{eq:PCSIlineI7}, $H(\brm{X}_{I})$, is lower bounded by $|I|L\geq 0$. Thus, in order to further lower bound $H(\brm{A}|\brm{Q},\brm{Y},\brm{X}_{\mathsf{W}})$, we need to lower bound the second term in~\eqref{eq:PCSIlineI7}, $H(\mathbf{A}|\mathbf{Q},\mathbf{Y},\mathbf{X}_{\mathsf{W}},\mathbf{Y}_I,\mathbf{X}_I)$. By the maximality of $I$, for each $j\in J\triangleq {\mathcal{K}\setminus (\mathsf{W}\cup \mathsf{S}\cup I)}$, there exists $\mathsf{C}_j\in \mathcal{C}$ (and $\mathbf{Y}_j\triangleq \mathbf{Y}^{[\mathsf{S},\mathsf{C}_j]}$, which is linearly dependent on $\mathbf{Y}$ and $\mathbf{Y}_I$) such that $H(\mathbf{X}_j|\mathbf{A},\mathbf{Q},\mathbf{Y}_j) = 0$, and as a consequence, $H(\mathbf{X}_j|\mathbf{A},\mathbf{Q},\mathbf{Y},\mathbf{Y}_I) = 0$. (Note that $|J|={K-M-1-|I|}$.) Let $\mathbf{X}_J\triangleq \{X_j\}_{j\in J}$. Then, we can write
\begin{align}
& H(\mathbf{A}|\mathbf{Q},\mathbf{Y},\mathbf{X}_{\mathsf{W}},\mathbf{Y}_I,\mathbf{X}_I)\nonumber\\ 
&	\quad = H(\mathbf{A}|\mathbf{Q},\mathbf{Y},\mathbf{X}_{\mathsf{W}},\mathbf{Y}_I,\mathbf{X}_I)\nonumber\\
& \quad\quad + H(\mathbf{X}_{J}|\mathbf{A},\mathbf{Q},\mathbf{Y},\mathbf{X}_{\mathsf{W}},\mathbf{Y}_{I},\mathbf{X}_I)\label{eq:PCSIlineI8}\\ 
& \quad = H(\brm{A},\brm{X}_{J}|\brm{Q},\brm{Y},\brm{X}_{\mathsf{W}},\brm{Y}_{I},\brm{X}_I)\nonumber\\
& \quad = H(\mathbf{X}_J|\mathbf{Q},\mathbf{Y},\mathbf{X}_{\mathsf{W}},\mathbf{Y}_I,\mathbf{X}_I)\nonumber\\
& \quad \quad + H(\mathbf{A}|\mathbf{Q},\mathbf{Y},\mathbf{X}_{\mathsf{W}},\mathbf{Y}_I,\mathbf{X}_I,\mathbf{X}_J)\nonumber\\
& \quad \geq H(\mathbf{X}_{J})\label{eq:PCSIlineI9}
\end{align} where~\eqref{eq:PCSIlineI8} holds since $H(\mathbf{X}_{j}|\mathbf{A},\mathbf{Q},\mathbf{Y},\mathbf{Y}_{I})=0$ for all $j\in J$ (by assumption); and~\eqref{eq:PCSIlineI9} holds because $\mathbf{X}_{J}$ and $(\mathbf{Q},\mathbf{Y},\mathbf{X}_{\mathsf{W}},\mathbf{Y}_I,\mathbf{X}_I)$ are independent by construction (noting that $J$ and ${\mathsf{W}\cup \mathsf{S}\cup I}$ are disjoint). Putting~\eqref{eq:PCSIlineI5},~\eqref{eq:PCSIlineI7}, and~\eqref{eq:PCSIlineI9} together, $H(\mathbf{A})\geq H(\mathbf{X}_{\mathsf{W}})+H(\mathbf{X}_I)+H(\mathbf{X}_{J}) = L+|I|L+ (K-M-1-|I|)L= (K-M)L$, as was to be shown.\QED

\subsection{Achievability}\label{subsec:AchThm1}
In this section, we propose a scalar-linear PIR-PCSI--I protocol that achieves the rate $(K-M)^{-1}$. The proposed protocol requires a base-field size $q\geq K$ (and arbitrary field-extension degree $l\geq 1$) where the messages $X_i$'s are elements from $\mathbb{F}_{q^{l}}$. 

It is noteworthy that the rate $(K-M)^{-1}$ is not necessarily achievable for ${q<K}$, and for the special case of scalar-linear schemes, the achievability of this rate is conditional upon the existence of a $(K,K-M)$ maximum distance seperable (MDS) code over $\mathbb{F}_q$ that has a codeword with support $\mathsf{W}\cup \mathsf{S}$ such that the $j$th code symbol is non-zero for $j=\mathsf{W}$ and it is equal to ${c}_j$ for each $j\in \mathsf{S}$ where ${c}_j$ is the coefficient of the message ${X}_j$ in the coded side information ${Y}^{[\mathsf{S},\mathsf{C}]}$. 

\vspace{0.25cm}
\textbf{Specialized GRS Code Protocol:} This protocol consists of three steps as follows:
 
\emph{Step~1:} First, the user arbitrarily chooses $K$ distinct elements $\omega_1,\dots,\omega_K$ from $\mathbb{F}_q$, and constructs a polynomial \[{p(x)\triangleq \prod_{i\in \mathcal{K}\setminus (\mathsf{W}\cup \mathsf{S})} (x-\omega_i)}.\] Then, the user constructs $K-M$ (ordered) sets ${Q}_1,\dots,{Q}_{K-M}$, each of size $K$, defined as \[{Q}_i=\{v_1\omega_1^{i-1},\dots,v_K\omega_K^{i-1}\},\] where the parameters $v_j$'s are chosen as follows. For each $j\in \mathsf{S}$, $v_j=\frac{c_j}{p(\omega_j)}$ where $c_j$ is the coefficient of $X_j$ in $Y^{[\mathsf{S},\mathsf{C}]}$; and for each $j\not\in \mathsf{S}$, $v_j$ is chosen at random from $\mathbb{F}_q^{\times}$. 

The user then sends to the server the query ${Q}^{[\mathsf{W},\mathsf{S},\mathsf{C}]} = \{{Q}_{1},\dots,{Q}_{K-M}\}$. 

Note that the $j$th element in the set ${Q}_i$ can be thought of as the entry $(i,j)$ of a $(K-M)\times K$ matrix $G\triangleq {[g_1^{\mathsf{T}},\dots,g_{K-M}^{\mathsf{T}}]}^{\mathsf{T}}$, which generates a $(K,K-M)$ GRS code with distinct parameters ${\omega_1,\dots,\omega_{K}}$ and non-zero multipliers $v_1,\dots,v_K$~\cite{Roth:06}. This construction ensures that such a GRS code has a specific codeword, namely $\sum_{i=1}^{K-M} p_{i} g_{i}$ where $p_i$ is the coefficient of $x^{i-1}$ in the expansion of the polynomial $p(x)= \sum_{i=1}^{K-M} p_{i}x^{i-1}$, with support $\mathsf{W}\cup \mathsf{S}$ such that the $j$th code symbol is non-zero for $j=\mathsf{W}$, and it is equal to $c_j$ for each $j\in \mathsf{S}$. This observation is the chief idea in the proof of the recoverability condition for the proposed protocol.

\emph{Step~2:} By using ${Q}_i$'s, the server computes $A_i$'s, defined as $A_{i} = \sum_{j=1}^{K}  v_j\omega_j^{i-1} X_{j}$, and it sends the answer ${A}^{[\mathsf{W},\mathsf{S},\mathsf{C}]}=\{A_{1},\dots,A_{K-M}\}$ to the user. 

Note that $A_i$'s are the parity check equations of a $(K,M)$ GRS code which is the dual code of the GRS code generated by the matrix $G$ defined earlier. 

\textbf{\it Step~3:} Upon receiving the answer, the user retrieves $X_{\mathsf{W}}$ by subtracting off the contribution of the side information $Y^{[\mathsf{S},\mathsf{C}]}$ from $\sum_{i=1}^{K-M} p_{i} A_{i} = c_{\mathsf{W}} X_{\mathsf{W}}+\sum_{i\in \mathsf{S}} c_{i}X_{i}$.

\begin{example}\label{ex:PCSI-I} 
Consider a scenario where the server has ${K=4}$ messages $X_1,\dots,X_{4}\in \mathbb{F}_{5}$, and the user demands the message $X_1$ and has a coded side information $Y = X_2+X_3$ with support size $M=2$. For this example, $\mathsf{W} = 1$, $\mathsf{S}= \{2,3\}$, and ${\mathsf{C} =\{c_2,c_3\}= \{1,1\}}$. 

First, the user chooses ${K=4}$ distinct elements $\omega_1,\dots,\omega_4$ from $\mathbb{F}_5$, say ${(\omega_1,\omega_2,\omega_3,\omega_4)=(0,1,2,3)}$. Then, the user constructs the polynomial \[p(x) = \prod_{i\not\in \mathsf{W}\cup \mathsf{S}} (x-\omega_i)={x-\omega_4}= x+2.\] Note that $p(x) = p_1+p_2x=2+x$. The user then computes $v_j$ for $j\in \mathsf{S}$, i.e., $v_2$ and $v_3$, by setting ${v_2=\frac{c_2}{p(\omega_2)}=2}$ and ${v_3=\frac{c_3}{p(\omega_3)}=4}$, and chooses $v_j$ for ${j\not\in \mathsf{S}}$, i.e., $v_1$ and $v_4$, at random (from $\mathbb{F}^{\times}_5$). Suppose that the user chooses $v_1=1$ and $v_4=2$. Then, the user constructs ${K-M=2}$ (ordered) sets ${Q}_1 = \{v_1,\dots,v_4\}=\{1,2,4,2\}$ and ${Q}_2 = \{v_1\omega_1,\dots,v_4\omega_4\}=\{0,2,3,1\}$. The user then sends the query $Q = \{Q_1,Q_2\}$ to the server. 

The server computes $A_1 = \sum_{j=1}^{4}v_j X_j = X_1+2X_2+4X_3+2X_4$ and $A_2 = \sum_{j=1}^{4}v_j\omega_j X_j = 2X_2+3X_3+X_4$, and sends the answer $A = \{A_1,A_2\}$ back to the user. Then, the user computes $\sum_{i=1}^{2} p_{i}A_i = 2A_1+A_2 = 2X_1+X_2+X_3$, and recovers $X_1$ by subtracting off $Y = X_2+X_3$. 

For this example, the rate of the proposed protocol is $1/2$. 

Note that the server knows the protocol, including the parameters $\omega_1,\dots,\omega_4$, and can compute the multipliers $v_1,\dots,v_4$, given the query. Since the side information coefficients $c_2$ and $c_3$ are uniformly distributed, the server finds each of the polynomials $x-\omega_1=x$, $x-\omega_2=4+x$, $x-\omega_3=3+x$, and $x-\omega_4=2+x$ equally likely to be the polynomial $p(x) = p_1+p_2x$, constructed in Step~1 of the protocol. Since the server knows that by the protocol the user requires the linear combination $p_1A_1+p_2A_2$ to recover the demand, from the server's perspective, each of the linear combinations $Z_1 = A_2$, $Z_2 = 4A_1+A_2$, $Z_3 = 3A_1+A_2$, $Z_4 = 2A_1+A_2$, i.e., $Z_1 = 2X_2+3X_3+X_4$, $Z_2 = 4X_1+4X_3+4X_4$, $Z_3 = 3X_1+3X_2+2X_4$, $Z_4 = 2X_1+X_2+X_3$, are equally likely to be the linear combination required by the user. Note, also, that, for each candidate demand index (e.g., $\{1\}$) and each candidate side information support index set (e.g., $\{2,3\}$), there exists exactly one of the linear combinations $Z_1,\dots,Z_4$ (e.g., $Z_4$) from which the candidate demand (e.g., $X_{1}$) can be recovered, given some linear combination (e.g., $X_2+X_3$) of the messages in the candidate side information support set (e.g., $X_2,X_3$). By these arguments, the server finds every index $i\in \{1,\dots,5\}$ and every pair of indices $\{i_1,i_2\}$ such that $i\not\in \{i_1,i_2\}$ equally likely to be the user's demand index and side information support index set, respectively. This confirms that the proposed protocol achieves $(\mathsf{W},\mathsf{S})$-privacy in this example. 
\end{example}

\begin{lemma}\label{lem:Ach1}
The Specialized GRS Code protocol is a scalar-linear PIR-PCSI--I protocol, and achieves the rate $(K-M)^{-1}$. 
\end{lemma}

\Proof
See Appendix~\ref{subsec:PLAch1}.\QED

\section{Proof of Theorem \ref{thm:PIRPCSI-II}}\label{sec:PIRPCSI-II}

\subsection{Converse}\label{subsec:ConvThm2}
First, we prove the converse for the scalar-linear case of Theorem~\ref{thm:PIRPCSI-II} for all $2\leq M\leq K$. The proof is based on a combination of algebraic and information-theoretic arguments. 

\begin{lemma}\label{lem:Conv2}
For any $2\leq M\leq K$, the scalar-linear capacity of the PIR-PCSI--II setting is upper bounded by ${(K-M+1)^{-1}}$.	
\end{lemma}

\Proof
In the following, all the entropies are conditional on the event $\mathds{1}_{\{\bsf{W}\in \bsf{S}\}}=1$, and for simplifying the notation, we remove this event from the conditions. We need to show that $H(\mathbf{A})\geq {(K-M+1)L}$. 

Let $I$ be the set of all $i\in \mathcal{K}$ such that $H(\mathbf{X}_i|\mathbf{A},\mathbf{Q})=0$. (Note that $0\leq |I|\leq K$). Let $\mathbf{X}_{I}\triangleq \{\mathbf{X}_i\}_{i\in I}$. By assumption, $\mathbf{X}_I$ and $\mathbf{Q}$ are independent and $H(\mathbf{X}_I|\mathbf{A},\mathbf{Q})=0$. Then, we can write 
\begin{align}
H(\mathbf{A}) & \geq H(\mathbf{A}|\mathbf{Q})\nonumber \\ 
&  = H(\mathbf{A}|\mathbf{Q})+H(\mathbf{X}_I|\mathbf{A},\mathbf{Q})\nonumber\\
& = H(\mathbf{A},\mathbf{X}_I|\mathbf{Q})\nonumber \\
& = H(\mathbf{X}_I|\mathbf{Q})+H(\mathbf{A}|\mathbf{Q},\mathbf{X}_I)\nonumber\\
& = H(\mathbf{X}_I)+H(\mathbf{A}|\mathbf{Q},\mathbf{X}_I).\label{eq:PCSIlineII10}
\end{align} If ${|I|\geq K-M+1}$, the first term in~\eqref{eq:PCSIlineII10}, $H(\brm{X}_I)$, is lower bounded by $(K-M+1)L$, and hence, $H(\mathbf{A})\geq {(K-M+1)L}$, as was to be shown. If $0\leq |I|\leq K-M$, the second term in~\eqref{eq:PCSIlineII10}, $H(\mathbf{A}|\mathbf{Q},\mathbf{X}_I)$, can be further lower bounded as follows. 

Assume, w.l.o.g., that $I=\{1,\dots,|I|\}$. (Note that $I = \emptyset$ for $|I|=0$.) Let $J\triangleq \{1,\dots,K-M-|I|+1\}$, and let $\mathsf{S}_j\triangleq {\{|I|+1,|I|+j+1,\dots,|I|+j+M-1\}}$ for $j\in J$. By Lemma~\ref{prop:1}, for each $j\in J$, there exists $\mathsf{C}_j\in \mathcal{C}$ (and $\mathbf{Y}_j\triangleq \mathbf{Y}^{[\mathsf{S}_j,\mathsf{C}_j]}$) such that $H(\mathbf{X}_{|I|+1}|\mathbf{A},\mathbf{Q},\mathbf{Y}_j)=0$. Let $\mathbf{Z}_j\triangleq \mathbf{Y}_j - c_j\mathbf{X}_{|I|+1}$ where $c_j$ is the coefficient of $\mathbf{X}_{|I|+1}$ in $\mathbf{Y}_j$. For any scalar-linear protocol where the answer consists only of scalar-linear combinations of messages in $X_{\mathcal{K}}$, it is easy to see that for each $j\in J$, (i) $H(\mathbf{Z}_j|\mathbf{A},\mathbf{Q})=0$, or (ii) ${H(\mathbf{Z}_j+c\mathbf{X}_{|I|+1}|\mathbf{A},\mathbf{Q})=0}$ for some $c\in \mathbb{F}^{\times}_q\setminus \{c_j\}$. (Otherwise, the server learns that $\mathsf{W}$ and $\mathsf{S}$ cannot be $|I|+1$ and $\mathsf{S}_j$, respectively. This obviously violates the $(\mathsf{W},\mathsf{S})$-privacy condition.) In either case (i) or (ii), one can see that $H(\mathbf{Z}_j|\mathbf{A},\mathbf{Q},\mathbf{X}_{|I|+1})=0$. (Note that this observation, which is the key in the proof of Lemma~\ref{lem:Conv2}, holds for all scalar-linear schemes, but not necessarily for all vector-linear or non-linear schemes in general. This implies the need for a different proof technique for the general schemes, and an example of such a technique is used in the proof of Lemma~\ref{lem:GConv2}.) Let $\mathbf{Z}_{J}\triangleq \{\mathbf{Z}_j\}_{j\in J}$. Then, we have
\begin{align}
H(\mathbf{A}|\mathbf{Q},\mathbf{X}_I) & \geq H(\mathbf{A}|\mathbf{Q},\mathbf{X}_I,\mathbf{X}_{|I|+1})\nonumber \\
& = 
H(\mathbf{A}|\mathbf{Q},\mathbf{X}_I,\mathbf{X}_{|I|+1})\nonumber\\
& \quad + H(\mathbf{Z}_J|\mathbf{A},\mathbf{Q},\mathbf{X}_I,\mathbf{X}_{|I|+1})\label{eq:PCSIlineII11}\\
& = H(\mathbf{A},\mathbf{Z}_J|\mathbf{Q},\mathbf{X}_I,\mathbf{X}_{|I|+1})\nonumber\\
& = H(\mathbf{Z}_J|\mathbf{Q},\mathbf{X}_I,\mathbf{X}_{|I|+1}) \nonumber \\
&\quad + H(\mathbf{A}|\mathbf{Q},\mathbf{X}_I,\mathbf{X}_{|I|+1},\mathbf{Z}_J)\nonumber\\
& \geq H(\mathbf{Z}_J) \label{eq:PCSIlineII12}
\end{align} where~\eqref{eq:PCSIlineII11} holds since $H(\mathbf{Z}_j|\mathbf{A},\mathbf{Q},\mathbf{X}_{|I|+1})=0$ for all $j\in J$ (by assumption); and~\eqref{eq:PCSIlineII12} follows because $\mathbf{Z}_J$ is independent of $(\mathbf{Q},\mathbf{X}_I,\mathbf{X}_{|I|+1})$ by construction, noting that $\mathbf{Z}_J$, $\mathbf{X}_I$, and $\mathbf{X}_{|I|+1}$ are linearly independent. By the linear independence of $\mathbf{Z}_j$'s for all $j\in J$, it follows that $H(\mathbf{Z}_J) = {(K-M-|I|+1)L}$. By~\eqref{eq:PCSIlineII10} and~\eqref{eq:PCSIlineII12}, we get $H(\mathbf{A})\geq H(\brm{X}_I) +H(\brm{Z}_J) = {|I|L}+{(K-M-|I|+1)L} ={(K-M+1)L}$, as was to be shown. \QED

Next, we give an information-theoretic proof of converse for the general case of Theorem~\ref{thm:PIRPCSI-II} for all ${\frac{K+1}{2}< M\leq K}$. For any $2\leq M\leq \frac{K+1}{2}$, the converse proof remains open. 

\begin{lemma}\label{lem:GConv2}
For any $\frac{K+1}{2}<M\leq K$, the capacity of the PIR-PCSI--II setting is upper bounded by ${(K-M+1)^{-1}}$. 
\end{lemma}

\Proof
Similar to the proof of Lemma~\ref{lem:Conv2}, for the ease of notation in the following we remove the event $\mathds{1}_{\{\bsf{W}\in \bsf{S}\}}=1$ from the conditions of all the entropies. We need to show that $H(\brm{A})\geq {(K-M+1)L}$. 

Let $J\triangleq \{1,\dots,K-M+1\}$ and $\mathsf{S}_j\triangleq \{j,\dots,{j+M-1}\}$ for $j\in J$. By Lemma~\ref{prop:1}, for each $j\in J$, there exists $\mathsf{C}_j\in \mathcal{C}$ (and $\brm{Y}_j\triangleq \brm{Y}^{[\mathsf{S}_j,\mathsf{C}_j]}$) such that $H(\brm{X}_{j}|\brm{A},\brm{Q},\brm{Y}_j)=0$. Let $\brm{X}_J\triangleq \{\brm{X}_j\}_{j\in J}$. (Note that $|J|={K-M+1}<M$ when $M>\frac{K+1}{2}$). Then, we can write
\begin{align} \label{eq:PCSIlineII13}
H(\brm{X}_J,\brm{Y}_J|\brm{Q}) & = H(\brm{X}_J,\brm{Y}_J) \\ \label{eq:PCSIlineII14}
& = 2(K-M+1)L,	
\end{align} where~\eqref{eq:PCSIlineII13} holds since $\brm{Q}$ is independent of $(\brm{X}_J,\brm{Y}_J)$ (by assumption); and~\eqref{eq:PCSIlineII14} follows because $\brm{X}_J$ and $\brm{Y}_J$ are independent by construction. (Note that $\brm{X}_J$ and $\brm{Y}_J$ are not necessarily independent for $|J| = K-M+1\geq M$, and a different technique---which remains open, is required for the proof of converse when $2\leq M\leq \frac{K+1}{2}$.) Moreover, we have
\begin{align}
H(\brm{X}_J,\brm{Y}_J|\brm{A},\brm{Q}) & \leq \sum_{j\in J}H(\brm{X}_j,\brm{Y}_j|\brm{A},\brm{Q}) \label{eq:PCSIlineII15}\\
&  = \sum_{j\in J} H(\brm{Y}_j|\brm{A},\brm{Q}) \nonumber \\
& \quad + \sum_{j\in J} H(\brm{X}_j|\brm{A},\brm{Q},\brm{Y}_j)\nonumber\\
&  = \sum_{j\in J} H(\brm{Y}_j|\brm{A},\brm{Q})\label{eq:PCSIlineII16}\\
& \leq \sum_{j\in J} H(\brm{Y}_j) \nonumber\\
&  = (K-M+1)L, \label{eq:PCSIlineII17} 	
\end{align} where~\eqref{eq:PCSIlineII15} follows from the chain rule of entropy;~\eqref{eq:PCSIlineII16} holds because $H(\brm{X}_{j}|\brm{A},\brm{Q},\brm{Y}_j)=0$ for $j\in J$ (by assumption); and~\eqref{eq:PCSIlineII17} holds because $\brm{Y}_j$'s for all $j\in J$ are independent by construction, and $\brm{Y}_j$ for each $j\in J$ is a scalar-linear combination of $\brm{X}_j,\brm{X}_{j+1},\dots,\brm{X}_{j+M-1}$. 

Using~\eqref{eq:PCSIlineII14} and~\eqref{eq:PCSIlineII17}, we can bound $H(\brm{X}_J,\brm{Y}_J,\brm{A}|\brm{Q})$ from below and above. On the one hand, we have
\begin{align}\nonumber
H(\brm{X}_J,\brm{Y}_J,\brm{A}|\brm{Q}) & 
\geq H(\brm{X}_J,\brm{Y}_J|\brm{Q}) \nonumber \\ 
& = 2(K-M+1)L, \label{eq:PCSIlineII18}
\end{align} where~\eqref{eq:PCSIlineII18} follows from~\eqref{eq:PCSIlineII14}. On the other hand, we have
\begin{align}
H(\brm{X}_J,\brm{Y}_J,\brm{A}|\brm{Q}) & = H(\brm{A}|\brm{Q}) \nonumber\\
& \quad +H(\brm{X}_J,\brm{Y}_J|\brm{A},\brm{Q}) \nonumber\\ 
& \leq H(\brm{A}|\brm{Q}) \nonumber \\
& \quad +(K-M+1)L, \label{eq:PCSIlineII19} 
\end{align} where~\eqref{eq:PCSIlineII19} follows from~\eqref{eq:PCSIlineII17}. Now, combining~\eqref{eq:PCSIlineII18} and~\eqref{eq:PCSIlineII19}, we have $H(\brm{A}|\brm{Q})\geq (K-M+1)L$, and as a consequence, $H(\brm{A})\geq H(\brm{A}|\brm{Q})\geq (K-M+1)L$, as was to be shown. \QED

\subsection{Achievability}\label{subsec:AchThm2}
In this section, we propose a scalar-linear PIR-PCSI--II protocol, termed the \emph{Modified Specialized GRS Code protocol}, that achieves the rate $(K-M+1)^{-1}$. For this protocol, the requirements for the parameters $q$ and $l$ are the same as those for the Specialized GRS Code protocol. 

\vspace{0.25cm}
\textbf{Modified Specialized GRS Code Protocol:} This protocol consists of three steps, where the steps 2-3 are the same as Steps 2-3 in the Specialized GRS Code protocol (Section~\ref{subsec:AchThm1}), when the parameter $M$ is replaced with $M-1$ everywhere. The step~1 of the proposed protocol is as follows: 

\emph{Step 1:} For $K$ arbitrarily chosen distinct elements $\omega_1,\dots,\omega_K$ from $\mathbb{F}_q$, the user constructs a polynomial \[{p(x) = \sum_{i=1}^{K-M+1} p_i x^{i-1} \triangleq \prod_{i\in\mathcal{K}\setminus \mathsf{S}} (x-\omega_i)},\] and constructs $K-M+1$ (ordered) sets $Q_1,\dots,Q_{K-M+1}$, each of size $K$, defined as \[Q_i=\{v_1\omega_1^{i-1},\dots,v_K\omega_K^{i-1}\},\] where $v_j$'s are chosen as follows. For each $j\in \mathsf{S}\setminus \mathsf{W}$, $v_j=\frac{c_j}{p(\omega_j)}$ where $c_j$ is the coefficient of $X_j$ in $Y^{[\mathsf{S},\mathsf{C}]}$; ${v_{\mathsf{W}}=\frac{c}{p(\omega_{\mathsf{W}})}}$ for a randomly chosen element $c$ from ${\mathbb{F}^{\times}_q\setminus \{c_{\mathsf{W}}\}}$ where $c_{\mathsf{W}}$ is the coefficient of $X_{\mathsf{W}}$ in $Y^{[\mathsf{S},\mathsf{C}]}$; and for each $j\not\in \mathsf{S}$, $v_j$ is chosen at random from $\mathbb{F}_q^{\times}$. 

The user then sends to the server the query ${Q}^{[\mathsf{W},\mathsf{S},\mathsf{C}]} = \{{Q}_{1},\dots,{Q}_{K-M+1}\}$.

\begin{example} Consider a scenario where the server has ${K=4}$ messages $X_1,\dots,X_{4}\in \mathbb{F}_{5}$, and the user demands the message $X_1$ and has a coded side information $Y = X_1+X_2$ with support size $M=2$. For this example, $\mathsf{W} = 1$, $\mathsf{S}= \{1,2\}$, and ${\mathsf{C} =\{c_1,c_2\}= \{1,1\}}$. 

First, the user chooses ${K=4}$ distinct elements $\omega_1,\dots,\omega_4$ from $\mathbb{F}_5$, say ${(\omega_1,\omega_2,\omega_3,\omega_4)=(0,1,2,3)}$. Then, the user constructs the polynomial \[p(x) = \prod_{i\not\in \mathsf{S}} (x-\omega_i)=(x-\omega_3)(x-\omega_4)= (x+3)(x+2).\] Note that $p(x) = p_1+p_2x+p_3x^2=1+x^2$. The user then computes $v_j$ for $j\in \mathsf{S}\setminus \mathsf{W}$, i.e., $v_2$, by setting ${v_2=\frac{c_2}{p(\omega_2)}=3}$; computes $v_{\mathsf{W}}$, i.e., $v_1$, for a randomly chosen element $c$, say $c = 4$, from $\mathbb{F}^{\times}_5\setminus \{c_1 = 1\}$ by setting $v_1 = \frac{c}{p(\omega_1)} = 4$; and chooses $v_j$ for ${j\not\in \mathsf{S}}$, i.e., $v_3$ and $v_4$, at random (from $\mathbb{F}^{\times}_5$). Suppose that the user chooses $v_3=1$ and $v_4=3$. Then, the user constructs ${K-M+1=3}$ (ordered) sets ${Q}_1 = \{v_1,\dots,v_4\}=\{4,3,1,3\}$, ${Q}_2 = \{v_1\omega_1,\dots,v_4\omega_4\}=\{0,3,2,4\}$, and ${Q}_3 = \{v_1\omega^2_1,\dots,v_4\omega^2_4\}=\{0,3,4,2\}$. The user then sends the query $Q = \{Q_1,Q_2,Q_3\}$ to the server. 

The server computes $A_1 = \sum_{j=1}^{4}v_j X_j = 4X_1+3X_2+X_3+3X_4$, $A_2 = \sum_{j=1}^{4}v_j\omega_j X_j = 3X_2+2X_3+4X_4$, and $A_3 = \sum_{j=1}^{4}v_j\omega^2_j X_j = 3X_2+4X_3+2X_4$, and sends the answer $A = \{A_1,A_2,A_3\}$ back to the user. Then, the user computes $\sum_{i=1}^{3} p_{i}A_i = A_1+A_3 = 4X_1+X_2$, and recovers $X_1$ by subtracting off $Y = X_1+X_2$. 

For this example, the rate of the proposed protocol is $1/3$. 

The proof of $(\mathsf{W},\mathsf{S})$-privacy for the proposed protocol in this example is similar to the proof of $(\mathsf{W},\mathsf{S})$-privacy for the Specialized GRS Code protocol in Example~\ref{ex:PCSI-I}.
\end{example}

\begin{lemma}\label{lem:Ach2}
The Modified Specialized GRS Code protocol is a scalar-linear PIR-PCSI--II protocol, and achieves the rate $(K-M+1)^{-1}$. 
\end{lemma}

\Proof
See Appendix~\ref{subsec:PLAch2}.
\QED

\section{Proof of Theorem \ref{thm:PIRCSI-I}}\label{sec:PIRCSI-I}

\subsection{Converse}\label{subsec:ConvThm3}
The capacity of the PIR-CSI--I setting is naturally upper bounded by the capacity of PIR with uncoded side information~\cite{Kadhe2017} where $M$ uncoded messages are available at the user as side information. As shown in~\cite{Kadhe2017}, the capacity of this problem is equal to $\lceil \frac{K}{M+1} \rceil^{-1}$, and the proof of this result relies on an index coding argument. In this section, we present an alternative converse proof for the case of general PIR-CSI--I protocols, by using information-theoretic arguments. Obviously, this proof also serves for the special case of scalar-linear PIR-CSI--I protocols.

\begin{lemma}\label{lem:Converse1}
For any $1\leq M\leq K-1$, the (scalar-linear) capacity of the PIR-CSI--I setting is upper bounded by ${\lceil \frac{K}{M+1}\rceil}^{-1}$.
\end{lemma}

\Proof 
In the following, all entropies are conditional on the event $\mathds{1}_{\{\bsf{W}\in \bsf{S}\}}=0$, and this event is removed from the conditions for the ease of notation. We need to show that $H(\brm{A})\geq \lceil\frac{K}{M+1}\rceil L$. 

Take arbitrary $\mathsf{W},\mathsf{S},\mathsf{C}$ (and $\brm{Y}\triangleq \brm{Y}^{[\mathsf{S},\mathsf{C}]}$) such that $\mathsf{W}\not\in \mathsf{S}$. Similar to the proof of Lemma~\ref{lem:Conv1}, it can be shown that 
\begin{equation}\label{eq:CSIlineI20}
H(\brm{A})\geq H(\brm{X}_{\mathsf{W}})+H(\brm{A}|\brm{Q},\brm{Y},\brm{X}_{\mathsf{W}}).	
\end{equation} 

There are two cases: (i) $W \cup S = \mathcal{K}$, and (ii) $W \cup S\neq \mathcal{K}$. In the case (i), $M = K-1$, and so, $\lceil \frac{K}{M+1}\rceil L = L$. Since $H(\brm{A}|\brm{Q},\brm{Y},\brm{X}_{\mathsf{W}})\geq 0$, then $H(\brm{A})\geq H(\brm{X}_{\mathsf{W}}) = L$ (by~\eqref{eq:CSIlineI20}), as was to be shown. In the case (ii), we proceed by lower bounding $H(\brm{A}|\brm{Q},\brm{Y},\brm{X}_{\mathsf{W}})$ as follows. 

We arbitrarily choose a message, say $\brm{X}_{\mathsf{W}_1}$, for some ${\mathsf{W}_1 \not\in \mathsf{W} \cup \mathsf{S}}$. By Lemma~\ref{prop:2}, there exist $\mathsf{S}_1\in \mathcal{S}$ with $\mathsf{W}_1\not\in \mathsf{S}_1$ and $\mathsf{C}_1\in \mathcal{C}$ (and $\brm{Y}_1 = \brm{Y}^{[\mathsf{S}_1,\mathsf{C}_1]}$) such that $H(\brm{X}_{\mathsf{W}_1}|\brm{A},\brm{Q},\brm{Y}_1) = 0$. Since conditioning does not increase the entropy, then $H(\brm{X}_{\mathsf{W}_1}|\brm{A},\brm{Q},\brm{Y},\brm{X}_{\mathsf{W}},\brm{Y}_1) = 0$. Thus, \begin{align}
H(\brm{A}|\brm{Q},\brm{Y},\brm{X}_{\mathsf{W}}) &\geq H\big(\brm{A}|\brm{Q},\brm{Y},\brm{X}_{\mathsf{W}},\brm{Y}_1\big)\nonumber\\
& = H(\brm{A}|\brm{Q},\brm{Y},\brm{X}_{\mathsf{W}},\brm{Y}_1) \nonumber\\ 
& \quad +H(\brm{X}_{\mathsf{W}_1}|\brm{A},\brm{Q},\brm{Y},\brm{X}_{\mathsf{W}},\brm{Y}_1)\nonumber \\
&= H\big(\brm{A},\brm{X}_{\mathsf{W}_1}|\brm{Q},\brm{Y},\brm{X}_{\mathsf{W}},\brm{Y}_1\big)\nonumber \\
&=H(\brm{X}_{\mathsf{W}_1}|\brm{Q},\brm{Y},\brm{X}_{\mathsf{W}},\brm{Y}_1)\nonumber \\ 
& \quad +H(\brm{A}|\brm{Q},\brm{Y},\brm{X}_{\mathsf{W}},\brm{Y}_1,\brm{X}_{\mathsf{W}_1})\nonumber\\ 
& = H(\brm{X}_{\mathsf{W}_1})\nonumber \\ & \quad +H(\brm{A}|\brm{Q},\brm{Y},\brm{X}_{\mathsf{W}},\brm{Y}_1,\brm{X}_{\mathsf{W}_1})\label{eq:CSIlineI21}
\end{align} where~\eqref{eq:CSIlineI21} holds because $\brm{X}_{\mathsf{W}_1}$ and $(\brm{Q},\brm{Y},\brm{X}_{\mathsf{W}},\brm{Y}_1)$ are independent (noting that ${\mathsf{W}_1\not\in \mathsf{W}\cup \mathsf{S}\cup \mathsf{S}_1}$), and hence, $H(\brm{X}_{\mathsf{W}_1}|\brm{Q},\brm{Y},\brm{X}_{\mathsf{W}},\brm{Y}_1) = H(\brm{X}_{\mathsf{W}_1})$.

Let $n \triangleq \lceil\frac{K}{M+1}\rceil$. Similarly as above, it can be shown that for all ${1\leq i\leq n-1}$ there exist $\mathsf{W}_1,\dots,\mathsf{W}_{i}\in \mathcal{K}$ and ${\mathsf{S}_1,\dots,\mathsf{S}_{i}\in \mathcal{S}}$ with $\mathsf{W}_j\not\in \mathsf{S}_j$ for all $1\leq j\leq i$ and ${\mathsf{W}_i\not\in \cup_{j=1}^{i-1} (\mathsf{W}_{j}\cup \mathsf{S}_{j}) \cup (\mathsf{W}\cup \mathsf{S})}$, and $\mathsf{C}_{1},\dots,\mathsf{C}_{i}\in \mathcal{C}$ (and $\brm{Y}_1 = \brm{Y}^{[\mathsf{S}_1,\mathsf{C}_1]},\dots,\brm{Y}_{i} = \brm{Y}^{[\mathsf{S}_i,\mathsf{C}_i]}$), such that \[H(\brm{X}_{\mathsf{W}_{i}}|\brm{A},\brm{Q},\brm{Y},\brm{X}_{\mathsf{W}},\brm{Y}_1,\brm{X}_{\mathsf{W}_1},\dots,\brm{Y}_{i-1},\brm{X}_{\mathsf{W}_{i-1}},\brm{Y}_{i})=0.\] Note that by construction, \[\left|\cup_{j=1}^{i-1} (\mathsf{W}_j \cup \mathsf{S}_j) \cup (\mathsf{W}\cup \mathsf{S})\right|\leq (M+1)i\] for all $1\leq i\leq n-1$. Repeating an argument similar to the one being used for lower bounding $H(\brm{A}|\brm{Q},\brm{Y},\brm{X}_{\mathsf{W}})$ as in~\eqref{eq:CSIlineI21}, it can be shown that
\begin{align*}
& H(\brm{A}|\brm{Q},\brm{Y},\brm{X}_{\mathsf{W}},\brm{Y}_1,\brm{X}_{\mathsf{W}_1},\dots,\brm{Y}_{i-1},\brm{X}_{\mathsf{W}_{i-1}})\nonumber \\ 
& \quad \geq H(\brm{X}_{\mathsf{W}_{i}}) +H(\brm{A}|\brm{Q},\brm{Y},\brm{X}_{\mathsf{W}},\brm{Y}_1,\brm{X}_{\mathsf{W}_1},\dots,\brm{Y}_{i},\brm{X}_{\mathsf{W}_{i}})
\end{align*} for all $1\leq i\leq n-1$. Combining these lower bounds for all $1\leq i\leq n-1$, we have
\begin{align}
H(\brm{A}|\brm{Q},\brm{Y},\brm{X}_{\mathsf{W}}) &\geq \sum_{i=1}^{n-1} H(\brm{X}_{\mathsf{W}_i})\nonumber \\
& = (n-1)L. \label{eq:CSIlineI22}
\end{align} Putting~\eqref{eq:CSIlineI20} and~\eqref{eq:CSIlineI22} together, we get $H(\brm{A})\geq nL = \lceil\frac{K}{M+1}\rceil L$. \QED

\subsection{Achievability}\label{subsec:AchThm3}
In this section, we propose a scalar-linear PIR-CSI--I protocol for arbitrary ${1\leq M\leq K-1}$. This protocol, termed \emph{Modified Partition-and-Code (MPC)}, is inspired by our recently proposed Partition-and-Code with Interference Alignment protocol in~\cite{HS2019} for private computation with uncoded side information. The MPC protocol does not make any assumption on the base-field size $q$ and the field-extension degree $l$, and is applicable for arbitrary $q\geq 2$ and $l\geq 1$. 

It should be noted that the Partition-and-Code protocol of~\cite{Kadhe2017} is only applicable to the PIR-CSI--I setting when $M+1$ divides $K$. Otherwise, when $M+1$ is not a divisor of $K$, the Partition-and-Code protocol will generate one part of size less than $M+1$. This immediately results in a violation of the $\mathsf{W}$-privacy condition. This is because the user's demand cannot be any of the messages pertaining to this part, noting that (i) the support set of the user's side information has size $M$, and (ii) all messages in the user's side information support set need to be combined with the user's demand.

\vspace{0.25cm}
\textbf{Modified Partition-and-Code (MPC) Protocol:} This protocol consists of three steps as follows: 

\emph{Step~1:} Let $n\triangleq\lceil \frac{K}{M+1}\rceil$. For $1\leq i\leq n-1$, we define $I_i \triangleq \{{(i-1)(M+1)+1},\dots,i(M+1)\}$, and $I_n\triangleq \{(n-1)(M+1)+1,\dots,K,1,\dots,n(M+1)-K\}$. (Note that $I_n = \{(n-1)(M+1)+1,\dots,K\}$ when $M+1$ divides $K$.) First, the user constructs a random permutation $\pi$ on $\mathcal{K} = \{1,\dots,K\}$ as follows.  

The user randomly chooses an index $j^{*}$ from $\mathcal{K}$, and assigns the demand index $\mathsf{W}$ to $\pi(j^{*})$, i.e., $\pi(j^{*}) = \mathsf{W}$. Let $i^{*}\triangleq \lceil \frac{j^{*}}{M+1}\rceil$ be the smallest index $i\in \{1,\dots,n\}$ such that $j^{*}\in I_i$. Then, the user randomly assigns the side information support indices in $\mathsf{S}$ to $\{\pi(j): j\in I_{i^{*}}\setminus \{j^{*}\}\}$ and randomly assigns the (not-yet-assigned) indices in $\mathcal{K}\setminus (\mathsf{W}\cup \mathsf{S})$ to $\{\pi(j): j\in \mathcal{K}\setminus I_{i^{*}}\}$. 

Next, the user constructs $n$ (ordered) sets $U_1,\dots,U_n$, each of size $M+1$, defined as $U_i = \{\pi(j): j\in I_i\}$; and constructs an (ordered) multiset $V$, defined as $V = \{c_{\pi(j)}: j\in I_{i^{*}}\}$ where $c_{\pi(j)}$ for $j\in I_{i^{*}}\setminus \{j^{*}\}$ is the coefficient of message $X_{\pi(j)}$ in the side information $Y^{[\mathsf{S},\mathsf{C}]}$, and $c_{\pi(j^{*})} = c_{\mathsf{W}}$ is a randomly chosen element from $\mathbb{F}^{\times}_q$. 

The user then constructs $Q_i = (U_i,V)$ for each ${1\leq i\leq n}$, and sends to the server the query $Q^{[\mathsf{W},\mathsf{S},\mathsf{C}]} = \{Q_{1},\dots,Q_{n}\}$. 

\textbf{\it Step~2:} By using $Q_i=(U_i,V)$'s, the server computes $A_i$'s, defined as $A_{i} = \sum_{j=1}^{M+1} c_{i_j} X_{i_j}$ where $U_{i} = \{i_1,\dots,i_{M+1}\}$ and $V = \{c_{i_1},\dots,c_{i_{M+1}}\}$, and sends back to the user the answer $A^{[\mathsf{W},\mathsf{S},\mathsf{C}]}=\{A_{1},\dots,A_{n}\}$.

\textbf{\it Step~3:} Upon receiving the answer from the server, the user retrieves $X_{\mathsf{W}}$ by subtracting off the contribution of the side information $Y^{[\mathsf{S},\mathsf{C}]}$ from $A_{i^{*}}=c_{\mathsf{W}}X_{\mathsf{W}}+\sum_{i\in \mathsf{S}} c_{i}X_{i}$.

\begin{example}
Consider a scenario where the server has ${K=5}$ messages $X_1,\dots,X_{5}\in \mathbb{F}_{3}$, and the user demands the message $X_1$ and has a coded side information $Y = X_2+2X_3$ with support size $M=2$. For this example, $\mathsf{W} = 1$, $\mathsf{S}= \{2,3\}$, and ${\mathsf{C} =\{c_2,c_3\}= \{1,2\}}$.

The parameters of the MPC protocol for this example are as follows: $n = \lceil\frac{K}{M+1}\rceil=2$, $I_1 = \{1,2,3\}$, and $I_2 = \{4,5,1\}$. 

First, the user constructs a permutation $\pi$ of $\{1,\dots,5\}$ as follows. The user randomly chooses an index $j^{*}$ from $\{1,\dots,5\}$, say $4$, and assigns the index $\mathsf{W} = 1$ to $\pi(j^{*}) = \pi(4)$, i.e., $\pi(4) = 1$. Note that, in this case, $i^{*}\triangleq \lceil \frac{j^{*}}{M+1}\rceil = 2$, and $I_{i^{*}} = I_2 = \{4,5,1\}$. The user then randomly assigns the indices in $\mathsf{S}$, i.e., $2$ and $3$, to $\{\pi(j): {j\in I_{i^{*}}\setminus \{j^{*}}\}\} = \{\pi(5),\pi(1)\}$, say $\pi(5)=3$ and $\pi(1)=2$; and randomly assigns the (not-yet-assigned) indices $4$ and $5$ to $\{\pi(j): j\in \{1,\dots,5\}\setminus I_{i^{*}}\} = \{\pi(2),\pi(3)\}$, say $\pi(2)=4$ and $\pi(3)=5$. Thus, the permutation $\pi$ maps $\{1,2,3,4,5\}$ to $\{2,4,5,1,3\}$. 

Next, the user constructs $n=2$ (ordered) sets $U_1,U_2$, each of size $M+1=3$, defined as $U_1 = \{\pi(j): j\in I_1\} = \{2,4,5\}$ and $U_2 = \{\pi(j): j\in I_2\} = \{1,3,2\}$; and constructs an (ordered) multiset $V$, defined as $V= \{c_{\pi(j)}: j\in I_{2}\} = \{c_1,c_3,c_2\}$ where $c_2=1$ and $c_3=2$ are the coefficients of $X_2$ and $X_3$ in the side information $Y$, and $c_1$ is a randomly chosen element from $\mathbb{F}^{\times}_3 = \{1,2\}$, say $c_1=2$. Thus, $V = \{2,2,1\}$. 

The user constructs $Q_1 = (U_1,V) = (\{2,4,5\},\{2,2,1\})$ and $Q_2 = (U_2,V) = (\{1,3,2\},\{2,2,1\})$, and sends the query $Q = \{Q_1,Q_2\}$ to the server. The server then computes $A_1 = 2X_2+2X_4+X_5$ and $A_2 = 2X_1+2X_3+X_2$, and sends the answer $A = \{A_1,A_2\}$ back to the user. Then, the user subtracts off the contribution of $Y = X_2+2X_3$ from $A_{i^{*}}=A_2 = 2X_2+X_2+2X_3$, and recovers $X_1$. 

For this example, the rate of the MPC protocol is $1/2$. Note that the rate of the Specialized GRS Code protocol---which achieves $(\mathsf{W},\mathsf{S})$-privacy and hence $\mathsf{W}$-privacy, for the scenario of this example is $(K-M)^{-1} = 1/3$.

From the perspective of the server, who knows the model and the parameters as well as the protocol, the messages $X_1,\dots,X_5$ are equally likely to be the user's demand. This is because, given the query, for each candidate demand, the server finds a unique potential side information. In particular, by the protocol, there must exist a linear combination $A_i$ in the answer $A = \{A_1,\dots,A_n\}$ (i.e., $\{A_1,A_2\}$ in this example) which is a function of the demand and the side information, and not a function of any other message. For example, given that the candidate demand is $X_1$, the server finds $X_2+2X_3$ as the only potential side information, noting that only $A_2 = 2X_1+X_2+2X_3$ is a linear combination of $X_1$ and $M=2$ other messages (i.e., $X_2$ and $X_3$). 

As an another example, consider the message $X_2$. Given that the candidate demand is $X_2$, there exist two linear combinations $A_1$ and $A_2$, each of which is a function of $X_2$ and $M=2$ other messages. However, by the protocol, among all linear combinations $A_i$ that are functions of the candidate demand and $M$ other messages, \emph{only the linear combination $A_i$ with the smallest index $i$ is a function of the demand and the side information.} Thus, for the candidate demand $X_2$, the server finds $2X_4+X_5$ as the only potential side information, noting that among $A_1$ and $A_2$---which are both functions of $X_2$ and $M=2$ other messages, the linear combination $A_1 = 2X_2+2X_4+X_5$ has the smallest index. Similarly, for each of the other candidate demands $X_3,X_4,X_5$, the server finds a unique potential side information. Moreover, the side information support index set is uniformly distributed and the demand index is conditionally distributed uniformly given the side information support index set. Putting these arguments together, one can see that given the query each message is equally likely to be the user's demand. This confirms that the MPC protocol satisfies the $\mathsf{W}$-privacy condition for this example. It is worth noting that the existence of a \emph{unique} potential side information for each candidate demand, which ensures $\mathsf{W}$-privacy, results in the violation of the $(\mathsf{W},\mathsf{S})$-privacy condition. For instance, in this example, given the query, for the candidate demand index $1$ the only potential side information support index set is $\{2,3\}$; and no other pair of indices in $\{2,\dots,5\}$ can be a potential side information support index set for the candidate demand index $1$.  
\end{example}

\begin{lemma}\label{lem:Ach3}
The Modified Partition-and-Code (MPC) protocol is a scalar-linear PIR-CSI--I protocol, and achieves the rate ${\lceil \frac{K}{M+1} \rceil}^{-1}$.
\end{lemma}

\Proof
See Appendix~\ref{subsec:PLAch3}.
\QED

\section{Proof of Theorem \ref{thm:PIRCSI-II}}\label{sec:PIRCSI-II}

\subsection{Converse}\label{subsec:ConvThm4}
In this section, we give an information-theoretic proof of converse for the case of general PIR-CSI--II protocols, which also serves as a converse proof for the special case of scalar-linear PIR-CSI--II protocols.

\begin{lemma}\label{lem:Converse2}
For $M=2$ and $M=K$, the (scalar-linear) capacity of the PIR-CSI--II setting is upper bounded by $1$, and for any $3\leq M\leq K-1$, the (scalar-linear) capacity of the PIR-CSI--II setting is upper bounded by $1/2$.
\end{lemma}

\Proof
In the following, all entropies are conditional on the event $\mathds{1}_{\{\bsf{W}\in \bsf{S}\}}=1$, and for simplifying the notation, we remove this event from the conditions everywhere. 

Take arbitrary $\mathsf{W},\mathsf{S},\mathsf{C}$ (and $\brm{Y}\triangleq \brm{Y}^{[\mathsf{S},\mathsf{C}]}$) such that $\mathsf{W}\in \mathsf{S}$. For the cases of $M=2$ and $M=K$, it suffices to show that $H(\brm{A})\geq L$. Note that $H(\brm{A})\geq H(\brm{A}|\brm{Q},\brm{Y})=H(\brm{A},\brm{X}_{\mathsf{W}}|\brm{Q},\brm{Y})$, where the equality follows from the recoverability condition, and $H(\brm{A},\brm{X}_{\mathsf{W}}|\brm{Q},\brm{Y})= H(\brm{X}_{\mathsf{W}}|\brm{Q},\brm{Y})+H(\brm{A}|\brm{Q},\brm{Y},\brm{X}_{\mathsf{W}})\geq H(\brm{X}_{\mathsf{W}})$, where the inequality follows from the independence of $\brm{X}_{\mathsf{W}}$ and $(\brm{Q},\brm{Y})$ by assumption. Putting these arguments together, $H(\brm{A})\geq H(\brm{X}_{\mathsf{W}})=L$.  

For the cases of $3\leq M \leq K-1$, we need to show that $H(\brm{A})\geq 2L$. By the above arguments, we have
\begin{equation}\label{eq:CSIlineII23}
H(\brm{A}) \geq H(\brm{X}_{\mathsf{W}})+H(\brm{A}|\brm{Q},\brm{Y},\brm{X}_{\mathsf{W}}).
\end{equation} 
Consider an arbitrary index $\mathsf{W}_1\in \mathsf{S}$. By the result of Lemma~\ref{prop:2}, there exist $\mathsf{S}_1\in \mathcal{S}$ with $\mathsf{W}_1\in \mathsf{S}_1$ and $\mathsf{C}_1\in \mathcal{C}$ (and $\brm{Y}_1 = \brm{Y}^{[\mathsf{S}_1,\mathsf{C}_1]}$) such that $H(\brm{X}_{\mathsf{W}_1}|\brm{A},\brm{Q},\brm{Y}_1) = 0$. Since conditioning does not increase the entropy, then $H(\brm{X}_{\mathsf{W}_1}|\brm{A},\brm{Q},\brm{Y},\brm{X}_{\mathsf{W}},\brm{Y}_1) = 0$. Then, we can write 
\begin{align}
H(\brm{A}|\brm{Q},\brm{Y},\brm{X}_{\mathsf{W}}) & \geq H(\brm{A}|\brm{Q},\brm{Y},\brm{X}_{\mathsf{W}},\brm{Y}_1) \nonumber \\
& = H(\brm{A}|\brm{Q},\brm{Y},\brm{X}_{\mathsf{W}},\brm{Y}_1) \nonumber \\
& \quad + H(\brm{X}_{\mathsf{W}_1}|\brm{A},\brm{Q},\brm{Y},\brm{X}_{\mathsf{W}},\brm{Y}_1) \nonumber\\
&=H(\brm{A},\brm{X}_{\mathsf{W}_1}|\brm{Q},\brm{Y},\brm{X}_{\mathsf{W}},\brm{Y}_1) \nonumber\\
&=H(\brm{X}_{\brm{W}_1}|\brm{Q},\brm{Y},\brm{X}_{\mathsf{W}},\brm{Y}_1)\nonumber \\&\quad + H(\brm{A}|\brm{Q},\brm{Y},\brm{X}_{\mathsf{W}},\brm{Y}_1,\brm{X}_{\mathsf{W}_1})\nonumber\\
& \geq H(\brm{X}_{\brm{W}_1}|\brm{Q},\brm{Y},\brm{X}_{\mathsf{W}},\brm{Y}_1). \label{eq:CSIlineII24}
\end{align} 

Noting that $\brm{Y},\brm{X}_{\mathsf{W}},\brm{Y}_1,\brm{X}_{\mathsf{W}_1}$ are linear functions of the messages in $\brm{X}_{\mathcal{K}}$, and $\brm{Q}$ is independent of $\brm{X}_{\mathcal{K}}$, there are two possible cases: (i) $H(\brm{X}_{\mathsf{W}_1}|\brm{Q},\brm{Y},\brm{X}_{\mathsf{W}},\brm{Y}_1)=H(\brm{X}_{\mathsf{W}_1})$, i.e., $\brm{X}_{\mathsf{W}_1}$ is independent of $(\brm{Q},\brm{Y},\brm{X}_{\mathsf{W}},\brm{Y}_1)$, or (ii) ${H(\brm{X}_{\mathsf{W}_1}|\brm{Q},\brm{Y},\brm{X}_{\mathsf{W}},\brm{Y}_1)=0}$, i.e., $\brm{X}_{\mathsf{W}_1}$ can be recovered from $\brm{Q},\brm{Y},\brm{X}_{\mathsf{W}},\brm{Y}_1$. 

In the case (i), $H(\brm{X}_{\mathsf{W}_1}|\brm{Q},\brm{Y},\brm{X}_{\mathsf{W}},\brm{Y}_1)=H(\brm{X}_{\mathsf{W}_1})$ by assumption. Rewriting~\eqref{eq:CSIlineII24}, we have
\begin{align}
H(\brm{A}|\brm{Q},\brm{Y},\brm{X}_{\mathsf{W}}) 
&\geq H(\brm{X}_{\mathsf{W}_1}).\label{eq:CSIlineII25} 
\end{align} By~\eqref{eq:CSIlineII23} and~\eqref{eq:CSIlineII25}, $H(\brm{A})\geq H(\brm{X}_{\mathsf{W}})+H(\brm{X}_{\mathsf{W}_1})=2L$.

In the case (ii), by assumption, $H(\brm{X}_{\mathsf{W}_1}|\brm{Q},\brm{Y},\brm{X}_{\mathsf{W}},\brm{Y}_1)=0$. Again, by the linearity of $\brm{Y},\brm{X}_{\mathsf{W}},\brm{Y}_1,\brm{X}_{\mathsf{W}_1}$, it must hold that $\brm{Y}=c_{\mathsf{W}} \brm{X}_{\mathsf{W}}+c_{\mathsf{W}_1} \brm{X}_{\mathsf{W}_1}+\brm{Z}$ and $\brm{Y}_1=c'_{\mathsf{W}} \brm{X}_{\mathsf{W}} + c'_{\mathsf{W}_1} \brm{X}_{\mathsf{W}_1}+c' \brm{Z}$ for some $c'_{\mathsf{W}},c'_{\mathsf{W}_1},c'\in \mathbb{F}^{\times}_q$, where $\brm{Z} = \sum_{i\in \mathsf{S}\setminus \{\mathsf{W},\mathsf{W}_1\}} c_{i} \brm{X}_i$. Unlike the previous case, this time we turn to an arbitrary index $\mathsf{W}_2\not\in \mathsf{S}$. Again, by the result of Lemma~\ref{prop:2}, there exist $\mathsf{S}_2\in \mathcal{S}$ with $\mathsf{W}_2\in \mathsf{S}_2$ and $\mathsf{C}_2\in \mathcal{C}$ (and $\brm{Y}_2 = \brm{Y}^{[\mathsf{S}_2,\mathsf{C}_2]}$) such that $H(\brm{X}_{\mathsf{W}_2}|\brm{A},\brm{Q},\brm{Y}_2) = 0$. Similar to~\eqref{eq:CSIlineII24}, it can be shown that 
\begin{align}
H(\brm{A}|\brm{Q},\brm{Y},\brm{X}_{\mathsf{W}}) & \geq H(\brm{X}_{\mathsf{W}_2}|\brm{Q},\brm{Y},\brm{X}_{\mathsf{W}},\brm{Y}_2)\nonumber \\&\quad + H(\brm{A}|\brm{Q},\brm{Y},\brm{X}_{\mathsf{W}},\brm{Y}_2,\brm{X}_{\mathsf{W}_2}). \label{eq:CSIlineII26}
\end{align} If $\brm{X}_{\mathsf{W}_2}$ is independent of $(\brm{Q},\brm{Y},\brm{X}_{\mathsf{W}},\brm{Y}_2)$, similarly as in the case (i) we can show that $H(\brm{A})\geq H(\brm{X}_{\mathsf{W}})+H(\brm{X}_{\mathsf{W}_2})=2L$. If $\brm{X}_{\mathsf{W}_2}$ is recoverable from $(\brm{Q},\brm{Y},\brm{X}_{\mathsf{W}},\brm{Y}_2)$, it must hold that $\brm{Y}_2=c''_{\mathsf{W}_2} \brm{X}_{\mathsf{W}_2}+c'' (c_{\mathsf{W}_1} \brm{X}_{\mathsf{W}_1}+\brm{Z})$ for some $c''_{\mathsf{W}_2},c''\in \mathbb{F}^{\times}_q$. Note that $\brm{X}_{\mathsf{W}_2}$ is independent of $(\brm{Q},\brm{Y}_1,\brm{X}_{\mathsf{W}_1},\brm{Y}_2)$ since by construction, $\brm{X}_{\mathsf{W}_2}$ cannot be recovered from $c'_{\mathsf{W}}\brm{X}_{\mathsf{W}}+c'\brm{Z}$ and $c''_{\mathsf{W}_2}\brm{X}_{\mathsf{W}_2}+c''\brm{Z}$, or in turn, from $\brm{Y}_1$ and $\brm{Y}_2$ given $\brm{X}_{\mathsf{W}_1}$. Also, $\brm{X}_{\mathsf{W}_1}$ is independent of $(\brm{Q},\brm{Y}_1,\brm{Y}_2)$ since, again by construction, $\brm{X}_{\mathsf{W}_1}$ cannot be recovered from $\brm{Y}_1$ and $\brm{Y}_2$. Thus, we can write
\begin{align}
H(\brm{A}) &\geq H(\brm{A}|\brm{Q},\brm{Y}_1,\brm{Y}_2)\nonumber \\ &=H(\brm{A},\brm{X}_{\mathsf{W}_1},\brm{X}_{\mathsf{W}_2}|\brm{Q},\brm{Y}_1,\brm{Y}_2) \label{eq:CSIlineII27}\\
& \geq H(\brm{X}_{\mathsf{W}_1}|\brm{Q},\brm{Y}_1,\brm{Y}_2)\nonumber\\
&\quad +H(\brm{X}_{\mathsf{W}_2}|\brm{Q},\brm{X}_{\mathsf{W}_1},\brm{Y}_1,\brm{Y}_2)\nonumber\\
& = H(\brm{X}_{\mathsf{W}_1})+H(\brm{X}_{\mathsf{W}_2})\label{eq:CSIlineII28}
\end{align} where~\eqref{eq:CSIlineII27} holds because $H(\brm{X}_{\mathsf{W}_1}|\brm{A},\brm{Q},\brm{Y}_1,\brm{Y}_2)=0$ and $H(\brm{X}_{\mathsf{W}_2}|\brm{A},\brm{Q},\brm{Y}_1,\brm{X}_{\mathsf{W}_1},\brm{Y}_2)=0$, noting that by assumption, ${H(\brm{X}_{\mathsf{W}_1}|\brm{A},\brm{Q},\brm{Y}_1)=0}$ and $H(\brm{X}_{\mathsf{W}_2}|\brm{A},\brm{Q},\brm{Y}_2)=0$; and~\eqref{eq:CSIlineII28} holds since as was shown earlier, $\brm{X}_{\mathsf{W}_1}$ and $\brm{X}_{\mathsf{W}_2}$ are independent of $(\brm{Q},\brm{Y}_1,\brm{Y}_2)$ and $(\brm{Q},\brm{Y}_1,\brm{X}_{\mathsf{W}_1},\brm{Y}_2)$, respectively. By~\eqref{eq:CSIlineII28}, we get $H(\brm{A})\geq 2L$.
\QED

\subsection{Achievability}\label{subsec:AchThm4}
In this section, we propose a scalar-linear PIR-CSI--II protocol for each of the following cases: (Case~1) $M=2$; (Case~2) $3\leq M\leq \frac{K}{2}+1$; (Case~3) $\frac{K+1}{2}\leq M\leq K-1$; and (Case~4) $M=K$. (Note that Cases~2 and~3 are overlapping at $M=\frac{K}{2}+1$ or $M=\frac{K+1}{2}$ when $K$ is even or odd, respectively. In these scenarios, either of the proposed protocols for Cases~2 and~3 applies.) It should be noted that the proposed protocols for Cases~1 and~2 are applicable for any base-field size $q\geq 2$ and any field-extension degree $l\geq 1$; whereas the proposed protocols for Cases~3 and~4 are applicable for any $q\geq 3$ and any $l\geq 1$. 

The proposed protocols are based on the idea of randomizing the \emph{structure} of query/answer, and are referred to as the \emph{Randomized Selection-and-Code (RSC) protocols}. In particular, in these protocols, for any given instance of the problem, there exist multiple different query/answer structures, each of which satisfies the recoverability condition; and one of these structures will be chosen at random according to a probability distribution, which is carefully designed to ensure the $\mathsf{W}$-privacy condition. 

For example, consider a scenario of Case~1 in which the server stores $X_1,X_2,X_3,\dots,X_K$, and the user's demand and side information are $X_1$ and $X_1+X_2$, respectively. The RSC protocol for Case~1 has two different (query/answer) structures: (i) the user queries $X_1$, which is the user's demand, and the server sends $X_1$ back to the user; or (ii) the user queries $X_2$, which is the other message in the user's side information, and the server sends back $X_2$ to the user. (Note that neither of these structures depend on the other messages $X_3,\dots,X_K$.) The RSC protocol for Case~1 randomly generates one of the two structures (i) and (ii), according to a probability distribution---specified shortly in the description of the protocol, designed in order to guarantee $\mathsf{W}$-privacy (i.e., given the query, each message in $X_1,\dots,X_K$ is equally likely to be the user's demand.) Note that using either of the two structures (i) and (ii), the user can recover $X_1$. The RSC protocols for Cases 2-4 use a similar idea.

For the ease of exposition, w.l.o.g., we assume that ${\mathsf{W}=\{1\}}$, ${\mathsf{S} = \{1,\dots,M\}}$, and ${\mathsf{C} = \{c_{1},\dots,c_{M}\}}$. 

\vspace{0.25cm}
\textbf{Randomized Selection-and-Code (RSC) Protocols:} The RSC protocol for each case consists of three steps, where the steps 2-3 are the same as Steps 2-3 in the MPC protocol (Section~\ref{subsec:AchThm3}). The step~1 of the RSC protocols are as follows: 

\vspace{0.25cm}
\textbf{Case 1:} The user randomly selects the index $\mathsf{W}$ (i.e., $1$) with probability $\frac{1}{K}$, or the other index in $\mathsf{S}$ (i.e., $2$) with probability $\frac{K-1}{K}$, and constructs two sets $U = \{i\}$ and $V=\{1\}$, where $i$ is the selected index by the user. 

The user then constructs $Q = (U,V)$, and sends the query $Q^{[\mathsf{W},\mathsf{S},\mathsf{C}]} = Q$ to the server. 

\begin{example}
Consider a scenario where the server has ${K=6}$ messages $X_1,\dots,X_{6}\in \mathbb{F}_{3}$, and the user demands the message $X_1$ and has a coded side information $Y = 2X_1+X_2$ with support size $M=2$. For this example, $\mathsf{W} = 1$, $\mathsf{S}= \{1,2\}$, and ${\mathsf{C} =\{c_1,c_2\}= \{2,1\}}$.

The user randomly selects an index $i$ from $\mathsf{S} = \{1,2\}$, where the probability of selecting the index $i=1$ is $\frac{1}{K} = \frac{1}{6}$, and the probability of selecting the index $i=2$ is $\frac{K-1}{K} = \frac{5}{6}$. Suppose that the user selects the index $i=2$. Then, the user requests the server for the message $X_i = X_2$, and the server responds by sending $X_2$ to the user. Subtracting off $X_2$ from $Y=2X_1+X_2$, the user then recovers $X_1$. 

For this example, the rate of the RSC protocol is $1$. Note that the Modified Specialized GRS Code protocol---which yields $(\mathsf{W},\mathsf{S})$-privacy and hence $\mathsf{W}$-privacy, achieves the rate ${(K-M+1)^{-1}} = 1/5$ for the scenario of this example.

From the server's perspective, the probability that the message $X_2$ is the user's demand is $\frac{1}{6}$, and the probability that one of the messages $X_1,X_3,\dots,X_6$ is the user's demand is $\frac{5}{6}$. Since these messages are equally likely to be the demand, the probability of any of them to be the user's demand is $\frac{5}{6}\times \frac{1}{5} = \frac{1}{6}$. This guarantees the $\mathsf{W}$-privacy.

Now, suppose that the user selects $i=1$. In this case, the user requests their demand $X_1$ from the server, and the server responds by sending $X_1$ back to the user. Again, from the perspective of the server, the probability that the message $X_1$ is the user's demand is $\frac{1}{6}$, and the probability of any of the messages $X_2,\dots,X_6$ to be the user's demand is $\frac{5}{6}\times\frac{1}{5} = \frac{1}{6}$. This again ensures the $\mathsf{W}$-privacy.
\end{example}

\textbf{Case 2:} The user constructs two (ordered) sets $U_1,U_2$, each of size $M-1$, with elements from the indices in $\mathcal{K}$, and an (ordered) multiset $V$ of size $M-1$ with elements from $\mathbb{F}^{\times}_q$. The constructions of $U_1,U_2,V$ are as follows. 

First, the user chooses an integer $r\in \{M-2,M-1\}$ by sampling from a probability distribution given by 
\begin{equation*}
\mathbb{P}(\brm{r}=r) = 
\begin{cases}
\frac{2M-2}{K}, & r = M-2,\\
1-\frac{2M-2}{K}, & r = M-1.
\end{cases}
\end{equation*} If $r=M-1$ is chosen, the user randomly selects $M-1$ indices from ${\mathcal{K}\setminus \mathsf{S}}$; otherwise, if $r=M-2$ is chosen, the user selects the index $\mathsf{W}$ along with $M-2$ randomly chosen indices from $\mathcal{K}\setminus \mathsf{S}$. Denote by $\{i_1,\dots,i_{M-1}\}$ the (ordered) set of the $M-1$ selected indices (in increasing order). Then, the user constructs $U_1 = \{2,\dots,M\}$ (i.e., the set of elements in $\mathsf{S}\setminus W$ in increasing order) and $U_2=\{i_1,\dots,i_{M-1}\}$. 

Next, the user constructs the (ordered) multiset $V=\{c_{2},\dots,c_{M}\}$ (i.e., the sequence of elements in $\mathsf{C}$ excluding the element $c_{\mathsf{W}}$). 

The user then constructs $Q_{i} = (U_{i},V)$ for each $i\in \{1,2\}$, and for a randomly chosen permutation $\sigma: \{1,2\}\mapsto \{1,2\}$, sends the query $Q^{[\mathsf{W},\mathsf{S},\mathsf{C}]} = \{Q_{\sigma(1)},Q_{\sigma(2)}\}$ to the server. 

\begin{example}
Consider a scenario where the server has ${K=6}$ messages $X_1,\dots,X_{6}\in \mathbb{F}_{3}$, and the user demands the message $X_1$ and has a coded side information $Y = 2X_1+X_2+2X_3$ with support size $M=3$. For this example, $\mathsf{W} = 1$, $\mathsf{S}= \{1,2,3\}$, and ${\mathsf{C} =\{c_1,c_2,c_3\}= \{2,1,2\}}$.

First, the user randomly chooses an integer $r\in \{M-2 = 1,M-1 = 2\}$, where the probability of choosing $r=1$ is $\frac{2}{3}$, and the probability of choosing $r=2$ is $\frac{1}{3}$. Suppose that the user chooses $r=1$. The user then selects the index $\mathsf{W}=1$ along with $r=1$ randomly chosen index from $\{1,\dots,6\}\setminus\{1,2,3\} = \{4,5,6\}$, say the index $4$. Then, the user constructs two (ordered) sets $U_1=\{2,3\}$ and $U_2 = \{1,4\}$, and the (ordered) multiset $V=\{c_2,c_3\}=\{1,2\}$.  

Then, the user constructs $Q_1 = (U_1,V) = (\{2,3\},\{1,2\})$ and $Q_2 = (U_2,V) = (\{1,4\},\{1,2\})$. For a randomly chosen permutation $\sigma$ on $\{1,2\}$, say $\sigma(1)=2$ and ${\sigma(2)=1}$, the user constructs the query $Q = \{Q_{\sigma(1)},Q_{\sigma(2)}\} = \{Q_2,Q_1\}$, and sends it to the server. The server computes $A_{i} = \sum_{j=1}^{M-1} c_{i_j} X_{i_j}$ for each $i\in \{1,2\}$ where $Q_i = (\{i_1,i_2\},\{c_{i_1},c_{i_2}\})$. For this example, $A_1 = X_2+2X_3$ and $A_2=X_1+2X_4$. Then, the server sends the answer $A=\{A_{\sigma(1)},A_{\sigma(2)}\}=\{A_2,A_1\}$ back to the user. Subtracting off $A_1$ from $Y=2X_1+X_2+2X_3$, the user recovers $X_1$. 

For this example, the rate of the RSC protocol is $1/2$; whereas the rate of the Modified Specialized GRS Code protocol for the scenario of this example is $(K-M+1)^{-1} = 1/4$. 

From the server's perspective, $U_1=\{2,3\}$ and $U_2=\{1,4\}$ are equally likely to be the index set of the user's side information support set (excluding the demand index). Let us refer to the event that $X_2$ and $X_3$ (or $X_1$ and $X_4$) are the two messages in the user's side information support set as E1 (or E2). Then, E1 (or E2) has probability $\frac{1}{2}$. Note also that, given E1 (or E2), $X_2$ and $X_3$ (or $X_1$ and $X_4$) have zero probability to be the user's demand.

Given E1, (i) with probability $\frac{1}{3}$, the user's demand is neither $X_1$ nor $X_4$, or (ii) with probability $\frac{2}{3}$, the user's demand is either $X_1$ or $X_4$. Given E1-(i), $X_5$ and $X_6$ are equally likely to be the user's demand. That is, given E1, $X_5$ (or $X_6$) is the user's demand with probability $\frac{1}{3}\times \frac{1}{2} = \frac{1}{6}$. Given E1-(ii), $X_1$ and $X_4$ are equally likely to be the user's demand. Then, given E1, $X_1$ (or $X_4$) is the user's demand with probability $\frac{2}{3}\times\frac{1}{2}=\frac{1}{3}$. 

Given E2, (i) with probability $\frac{1}{3}$, the user's demand is neither $X_2$ nor $X_3$, or (ii) with probability $\frac{2}{3}$, the user's demand is either $X_2$ or $X_3$. Given E2-(i), either of $X_5$ and $X_6$ is the user's demand with probability $\frac{1}{2}$. Then, given E2, $X_5$ (or $X_6$) is the user's demand with probability $\frac{1}{3}\times \frac{1}{2} = \frac{1}{6}$. Given E2-(ii), either of $X_2$ and $X_3$ is the user's demand with probability $\frac{1}{2}$. Then, given E1, $X_2$ (or $X_3$) is the user's demand with probability $\frac{2}{3}\times\frac{1}{2}=\frac{1}{3}$. 

From the above arguments, it is easy to see that given the query, each message $X_i$ is equally likely to be the user's demand, and hence the $\mathsf{W}$-privacy condition is satisfied. For example, $X_1$ has probability $\frac{1}{3}$ (or $0$) to be the user's demand given E1 (or E2). Since E1 and E2 each have probability $\frac{1}{2}$, the probability of $X_1$ to be the user's demand is $\frac{1}{2}\times\frac{1}{3}+\frac{1}{2}\times 0 = \frac{1}{6}$. As an another example, consider $X_5$. Given either of E1 or E2, $X_5$ has probability $\frac{1}{6}$ to be the user's demand. Thus, the probability of $X_5$ to be the user's demand is $\frac{1}{2}\times\frac{1}{6}+\frac{1}{2}\times \frac{1}{6} = \frac{1}{6}$.   
\end{example}

\textbf{Case 3:} The user constructs two (ordered) sets $U_1,U_2$, each of size $M$, with elements from the indices in $\mathcal{K}$, and an (ordered) multiset $V$ of size $M$ with elements from $\mathbb{F}^{\times}_q$. The constructions of $U_1,U_2,V$ are as follows. 

The user chooses an integer ${s\in\{2M-K-1,2M-K\}}$ by sampling from a probability distribution given by
\begin{equation*}
\mathbb{P}(\brm{s}=s) = 
\begin{cases}
1-\frac{2K-2M}{K}, & s = 2M-K-1,\\
\frac{2K-2M}{K}, & s = 2M-K.
\end{cases}
\end{equation*} If $s=2M-K$ is chosen, the user randomly selects $2M-K$ indices from $\mathsf{S}\setminus \mathsf{W}$; otherwise, if $s=2M-K-1$ is chosen, the user selects the index $\mathsf{W}$ together with $2M-K-1$ randomly chosen indices from $\mathsf{S}\setminus \mathsf{W}$. Denote by $\{i_1,\dots,i_{M}\}$ the (ordered) set of the $2M-K$ selected indices and the $K-M$ indices in $\mathcal{K}\setminus \mathsf{S}$ (in increasing order). Then, the user constructs ${U_1 = \{1,\dots,M\}}$ (i.e., the set of elements in $\mathsf{S}$ in increasing order) and $U_2=\{i_1,\dots,i_M\}$.

Next, the user constructs the (ordered) multiset $V=\{c,c_{2},\dots,c_{M}\}$ (i.e., the sequence of the elements in $\mathsf{C}$, except when the element $c_{\mathsf{W}}$ is replaced by the element $c$) where $c$ is randomly chosen from ${\mathbb{F}^{\times}_q\setminus \{c_{1}\}}$ (i.e., $\mathbb{F}^{\times}_q\setminus \{c_{\mathsf{W}}\}$). 

The user then constructs $Q_i = (U_i,V)$ for each $i\in \{1,2\}$, and for a randomly chosen permutation $\sigma: \{1,2\}\mapsto \{1,2\}$, sends the query $Q^{[\mathsf{W},\mathsf{S},\mathsf{C}]} = \{Q_{\sigma(1)},Q_{\sigma(2)}\}$ to the server. 

\vspace{0.25cm}
\textbf{Case 4:} The user creates two (ordered) sets $U = \{1,\dots,K\}$ and $V=\{c,c_{2},\dots,c_{K}\}$ (i.e., the sequence of elements in $\mathsf{C}$, except when the element $c_{\mathsf{W}}$ is replaced by the element $c$) where $c$ is randomly chosen from $\mathbb{F}^{\times}_q\setminus \{c_{1}\}$ (i.e., $\mathbb{F}^{\times}_q\setminus \{c_{\mathsf{W}}\}$). 

The user then constructs $Q = (U,V)$, and sends the query $Q^{[\mathsf{W},\mathsf{S},\mathsf{C}]} = Q$ to the server. 

\begin{lemma}\label{lem:Ach4}
The Randomized Selection-and-Code (RSC) protocols for $M=2$, $3\leq M\leq \frac{K}{2}+1$, ${\frac{K+1}{2}\leq M\leq K-1}$, and $M=K$ are scalar-linear PIR-CSI--II protocols, and achieve the rates $1$, $1/2$, $1/2$, and $1$, respectively.
\end{lemma}

\Proof
See Appendix~\ref{subsec:PLAch4}.
\QED

\section{Conclusion and Future Work}
In this work, we studied the fundamental limits of single-message single-server information-theoretic PIR in the presence of a coded side information. Considering two different types of privacy, namely $(\mathsf{W},\mathsf{S})$-privacy and $\mathsf{W}$-privacy, we characterized the capacity and the scalar-linear capacity of the problem under two different models depending on whether the support set of the user's coded side information includes the requested message or not. In addition, for each problem setting we proposed a novel scalar-linear scheme that achieves the capacity. 

One natural question that remains open is that how much the capacity will increase if we relax the assumption that the server knows the considered model, i.e., whether the side information is a function of the demand or not. Our preliminary results, beyond the scope of this work and hence not presented here, suggest that in an asymptotic regime (when the number of messages in the database grows unbounded), the capacity remains the same even if the server is aware of whether the side information depends on the demand or not. A detailed study of this observation remains open. 

Another direction for future work is to characterize the capacity of the single-server PIR when the user has multiple coded side information and/or wants multiple messages from the server. Our initial attempts at studying these settings suggest that there is a close relation between these problems and the problem of single-server private computation with coded side information, which is the focus of an ongoing work. 

Last but not least, the extensions of this work for multi-server setting were also recently studied in~\cite{KKHS12019} and~\cite{KKHS22019}  for the cases in which $\mathsf{W}$-privacy and $(\mathsf{W},\mathsf{S})$-privacy are required, respectively. Some achievability schemes based on those in this work were proposed; notwithstanding, the capacity of these settings are still open in general.

\appendices

\section{Proofs of Lemmas~\ref{lem:Ach1} and~\ref{lem:Ach2}}\label{sec:A1}

\subsection{Proof of Lemma~\ref{lem:Ach1}}\label{subsec:PLAch1}
Since the matrix $G$, defined in Step~1 of the Specialized GRS Code protocol, generates a $(K,K-M)$ GRS code which is an MDS code, the rows of $G$ are linearly independent. Accordingly, $\brm{A}_1,\dots,\brm{A}_{K-M}$, defined in Step~2, are linearly independent combinations of the messages in $\brm{X}_{\mathcal{K}}$, which are themselves independently and uniformly distributed over $\mathbb{F}_{q^{l}}$. This implies that $\brm{A}_1,\dots,\brm{A}_{K-M}$ are independently and uniformly distributed over $\mathbb{F}_{q^{l}}$. Since $H(\mathbf{X}_j)=L$ for all $j\in \mathcal{K}$, then $H(\mathbf{A}_i)=L$ for all $i\in \{1,\dots,K-M\}$. Thus, for all $\mathsf{W}\in \mathcal{K},\mathsf{S}\in \mathcal{S},\mathsf{C}\in \mathcal{C}$ such that $\mathsf{W}\not\in\mathsf{S}$, we have $H(\mathbf{A}^{[\mathsf{W},\mathsf{S},\mathsf{C}]}) = H(\mathbf{A}_1,\dots,\mathbf{A}_{K-M})=\sum_{i=1}^{K-M} H(\mathbf{A}_i)=(K-M)L$. (Note that $H(\mathbf{A}^{[\mathsf{W},\mathsf{S},\mathsf{C}]})=(K-M)L$ does not depend on the realizations $\mathsf{W},\mathsf{S},\mathsf{C}$.) Given that ${\bsf{W}\not\in \bsf{S}}$, $\bsf{W}$ and $\bsf{S}$ are jointly distributed uniformly, and $\bsf{C}$ is distributed uniformly (and independently from $(\bsf{W},\bsf{S})$). Thus, $H(\mathbf{A}^{[\bsf{W},\bsf{S},\bsf{C}]}|\bsf{W}\not\in \bsf{S})=H(\mathbf{A}^{[\mathsf{W},\mathsf{S},\mathsf{C}]}) = (K-M)L$, implying that the rate of the Specialized GRS Code protocol is equal to ${L/((K-M)L)} = (K-M)^{-1}$. 

The scalar-linearity of $\brm{A}_i$'s in the messages $\brm{X}_{j}$'s confirms that the Specialized GRS Code protocol is scalar-linear. From the construction, it should also be obvious that the recoverability condition is satisfied. The proof of $(\mathsf{W},\mathsf{S})$-privacy relies on two facts: (i) the $(K,K-M)$ GRS code, generated by the matrix $G$, is an MDS code, and hence the minimum (Hamming) weight of a codeword is $K-(K-M)+1 = M+1$; and (ii) there exist the same number of minimum-weight codewords for any support of size ${M+1}$~\cite{Roth:06}. The rest of the proof is as follows. 

From (i) and (ii), for any $\mathsf{W}^{*}\in \mathcal{K}, \mathsf{S}^{*}\in \mathcal{S}$ such that ${\mathsf{W}^{*}\not\in \mathsf{S}^{*}}$, the dual code, whose parity check matrix is $G$, contains the same number of parity check equations with support $\mathsf{W}^{*}\cup \mathsf{S}^{*}$ (i.e., the messages $\{X_i\}_{i\in \mathsf{W}^{*}\cup \mathsf{S}^{*}}$ have non-zero coefficients and the rest of the messages all have zero coefficients). For given $\mathsf{W}^{*},\mathsf{S}^{*}$, consider an arbitrary such parity check equation $Z = c_{\mathsf{W}^{*}}X_{\mathsf{W}^{*}}+\sum_{i\in \mathsf{S}^{*}}  c_i X_i$ where $c_{i}\in \mathbb{F}^{\times}_q$ for all $i\in \mathsf{W}^{*}\cup \mathsf{S}^{*}$. The candidate demand $X_{\mathsf{W}^{*}}$ can be recovered from $Z$, only given a potential side information $\sum_{i\in \mathsf{S}^{*}} c (c_i X_i)$ for arbitrary $c\in \mathbb{F}^{\times}_q$. Noting that $|\mathbb{F}^{\times}_q| = q-1$, for any given parity check equation $Z$ with support $\mathsf{W}^{*}\cup \mathsf{S}^{*}$, there exist only $q-1$ potential side information, namely $\{c(Z-\sum_{i\in \mathsf{S}^{*}} c_i X_i): c\in \mathbb{F}^{\times}_q\}$, from each of which the candidate demand $X_{\mathsf{W}^{*}}$ can be recovered. This proves the $(\mathsf{W},\mathsf{S})$-privacy of the Specialized GRS Code protocol. 

\subsection{Proof of Lemma~\ref{lem:Ach2}}\label{subsec:PLAch2}
The proof, omitted to avoid repetition, follows from the same lines as in the proof of Lemma~\ref{lem:Ach1} (Appendix~\ref{subsec:PLAch1}). 

\section{Proofs of Lemmas~\ref{lem:Ach3} and~\ref{lem:Ach4}}\label{sec:A2}

\subsection{Proof of Lemma~\ref{lem:Ach3}}\label{subsec:PLAch3}
By the construction of the Modified Partition-and-Code (MPC) protocol (see Steps~1-2), $\brm{A}_1,\dots,\brm{A}_n$ are linearly independent combinations of the messages in $\brm{X}_{\mathcal{K}}$. Using a similar argument as the one in the proof of Lemma~\ref{lem:Ach1} (Appendix~\ref{subsec:PLAch1}), it can be shown that $H(\brm{A}^{[\mathsf{W},\mathsf{S},\mathsf{C}]}) = H(\brm{A}_1,\dots,\brm{A}_n) = nL$ for all $\mathsf{W}\in \mathcal{K}, \mathsf{S}\in \mathcal{S},\mathsf{C}\in \mathcal{C}$ such that $\mathsf{W}\not\in \mathsf{S}$, and $H(\brm{A}^{[\bsf{W},\bsf{S},\bsf{C}]}|\bsf{W}\not\in \bsf{S})= H(\brm{A}^{[\mathsf{W},\mathsf{S},\mathsf{C}]}) = nL$. This implies that the rate of the MPC protocol is equal to $L/nL = \lceil\frac{K}{M+1}\rceil^{-1}$. 

The scalar-linearity of the MPC protocol should be obvious from the construction. The recoverability condition is also obviously satisfied (see Step~3). 

To prove that the MPC protocol satisfies the $\mathsf{W}$-privacy condition, we need to show that for any query $Q$ generated by the protocol, \[{\mathbb{P}(\bsf{W}=\mathsf{W}|\mathbf{Q}=Q,\bsf{W}\not\in \bsf{S})=\mathbb{P}(\bsf{W}=\mathsf{W}|\bsf{W}\not\in \bsf{S})}\] for all ${\mathsf{W}\in \mathcal{K}}$, or in turn, $\mathbb{P}(\bsf{W}=\mathsf{W}|\mathbf{Q}=Q,\bsf{W}\not\in \bsf{S})$ does not depend on $\mathsf{W}$. (Note that by construction, $\brm{Q}$ is independent of the messages in $\brm{X}_{\mathcal{K}}$.)

By Step~1 of the protocol, for any given $\mathsf{W}\in \mathcal{K}$, there exist a unique $\mathsf{S}_{\mathsf{W}}\in \mathcal{S}$ (with $\mathsf{W}\not\in \mathsf{S}_{\mathsf{W}}$) and a unique $\mathsf{C}_{\mathsf{W}}\in \mathcal{C}$ such that the triple $(\mathsf{W},\mathsf{S}_{\mathsf{W}},\mathsf{C}_{\mathsf{W}})$ complies with the query $Q$, i.e., given that $X_{\mathsf{W}}$ and $Y^{[\mathsf{S}_{\mathsf{W}},\mathsf{C}_{\mathsf{W}}]}$ are the user's demand and side information, respectively, the protocol could potentially generate the query $Q$. Then, we have \begin{align*}
&\mathbb{P}(\bsf{W}=\mathsf{W}|\mathbf{Q}=Q, \bsf{W}\not\in \bsf{S})  \\ 
& = {\mathbb{P}(\bsf{W}=\mathsf{W},\bsf{S}=\mathsf{S}_{\mathsf{W}},\bsf{C}=\mathsf{C}_{\mathsf{W}}|\mathbf{Q}=Q,\bsf{W}\not\in \bsf{S})}.
\end{align*} Since the conditional distribution of $(\bsf{W},\bsf{S},\bsf{C})$ given $\bsf{W}\not\in \bsf{S}$ is uniform, by applying the Bayes' rule one can see that ${\mathbb{P}(\bsf{W}=\mathsf{W},\bsf{S}=\mathsf{S}_{\mathsf{W}},\bsf{C}=\mathsf{C}_{\mathsf{W}}|\mathbf{Q}=Q,\bsf{W}\not\in \bsf{S})}$ does not depend on $\mathsf{W}$ so long as ${\mathbb{P}(\mathbf{Q}=Q|\bsf{W}=\mathsf{W},\bsf{S}=\mathsf{S}_{\mathsf{W}},\bsf{C}=\mathsf{C}_{\mathsf{W}})}$ does not depend on $\mathsf{W}$. By the design of the protocol, 
\begin{align*}
& \mathbb{P}(\mathbf{Q}=Q|\bsf{W}=\mathsf{W},\bsf{S}=\mathsf{S}_{\mathsf{W}},\bsf{C}=\mathsf{C}_{\mathsf{W}}) \\ 
& =\frac{1}{K!}\binom{K-1}{M}(q-1)^{-1}	
\end{align*} for all $\mathsf{W}\in \mathcal{K}$, and hence $\mathbb{P}(\bsf{W}=\mathsf{W}|\mathbf{Q} = Q,\bsf{W}\not\in \bsf{S})$ does not depend on $\mathsf{W}$.

\subsection{Proof of Lemma~\ref{lem:Ach4}}\label{subsec:PLAch4}
The proofs for the rates of the RSC protocols follow the same line as in the proof of the rate of the MPC protocol in Lemma~\ref{lem:Ach3} (Appendix~\ref{subsec:PLAch3}), and hence omitted. From the construction, it should also be obvious that the RSC protocols are scalar-linear. Moreover, it should not be hard to see from the description of these protocols that the recoverability condition is satisfied.

To prove that the RSC protocols satisfy the $\mathsf{W}$-privacy condition, we need to show that \[{\mathbb{P}(\bsf{W}=\mathsf{W}|\brm{Q}=Q,\bsf{W}\in \bsf{S})}={\mathbb{P}(\bsf{W}=\mathsf{W}|\bsf{W}\in \bsf{S})}\] for all $\mathsf{W}\in \mathcal{K}$. Alternatively, by the Bayes' rule, it suffices to show that ${\mathbb{P}(\brm{Q}=Q|\bsf{W}=\mathsf{W},\bsf{W}\in\bsf{S})}$ does not depend on $\mathsf{W}$.

Recall that $Q = (U,V)$ for Cases~1 and~4, and $Q = \{Q_1,Q_2\} = \{(U_1,V),(U_2,V)\}$ for Cases~2 and~3. For simplifying the notation, let us denote $\{U_1,U_2\}$ by $U$ for Cases~2 and~3. By the construction of the RSC protocols and the model assumptions, given $\bsf{W}\in \bsf{S}$, the following two observations hold: (i) $\brm{U}$ and $\bsf{V}$ are conditionally independent given $\bsf{W}$, and (ii) $\brm{V}$ and $\bsf{W}$ are independent. The observation (i) should be obvious, and the observation (ii) holds because $\bsf{V}$ is uniformly distributed over all possible choices of $V$ for each case. (For example, for Case~1, $\brm{V}=\{1\}$; and for Case~2, $\brm{V}=\{\brm{c}_i: i\in \bsf{S}\setminus \bsf{W}\}$---where $\brm{c}_i$'s are uniformly distributed over $\mathbb{F}^{\times}_q$, is uniformly distributed over all ordered multisets of size $M-1$ with elements from $\mathbb{F}^{\times}_q$.) Using~(i) and~(ii),
\begin{align*}
& \mathbb{P}(\brm{Q}=Q|\bsf{W}=\mathsf{W},\bsf{W}\in\bsf{S})	 \\
& = \mathbb{P}(\brm{U}=U, \brm{V}=V|\bsf{W}=\mathsf{W},\bsf{W}\in\bsf{S})\\
& = \mathbb{P}(\brm{V}=V|\bsf{W}\in\bsf{S})\times 	\mathbb{P}(\brm{U}=U|\bsf{W}=\mathsf{W},\bsf{W}\in\bsf{S}).
\end{align*} Since $\mathbb{P}(\brm{V}=V|\bsf{W}\in\bsf{S})$ does not depend on $\mathsf{W}$, instead of showing that ${\mathbb{P}(\brm{Q}=Q|\bsf{W}=\mathsf{W},\bsf{W}\in\bsf{S})}$ is not a function of $\mathsf{W}$, it suffices to show that
${\mathbb{P}(\brm{U}=U|\bsf{W}=\mathsf{W},\bsf{W}\in\bsf{S})}$ does not depend on $\mathsf{W}$. In the following, we prove this claim for the RSC protocol for each case separately.  

With a slight abuse of notation, hereafter for the ease of exposition, we treat the ordered sets $U_1,U_2$ as (unordered) sets.\\  

\subsubsection*{Case~1} For an arbitrary $i\in \mathcal{K}$, consider $U=\{i\}$. Take an arbitrary $\mathsf{W}\in \mathcal{K}$. There are two cases as follows: (i) $\mathsf{W}=i$, and (ii) $\mathsf{W}\neq i$. 

In the case (i), we have  
\begin{align}
& \mathbb{P}(\brm{U}=U|\bsf{W}=\mathsf{W},\bsf{W}\in \bsf{S}) \nonumber\\ 
&  =\sum_{j\in\mathcal{K}\setminus \mathsf{W}}\mathbb{P}(\brm{U}=U|\bsf{W}=\mathsf{W}, \bsf{S}=\{\mathsf{W},j\})\nonumber\\
& \quad\quad\quad\hspace{0.275cm} \times \mathbb{P}(\bsf{S}=\{\mathsf{W},j\}|\bsf{W}=\mathsf{W},\bsf{W}\in \bsf{S}).\label{eq:CSIlineII29}
\end{align} By the model assumption, we have
\begin{equation}\label{eq:CSIlineII30}
\mathbb{P}(\bsf{S}=\{\mathsf{W},j\}|\bsf{W}=\mathsf{W},\bsf{W}\in \bsf{S})=\frac{1}{K-1}	
\end{equation} for all $j\in \mathcal{K}\setminus \mathsf{W}$. Moreover, given that $\bsf{W}=\mathsf{W}$ and ${\bsf{S}=\{\mathsf{W},j\}}$, the protocol constructs $U = \{\mathsf{W}\}$ with probability $\frac{1}{K}$. This implies that
\begin{equation}\label{eq:CSIlineII31}
\mathbb{P}(\brm{U}=U|\bsf{W}=\mathsf{W}, \bsf{S}=\{\mathsf{W},j\})=\frac{1}{K}	
\end{equation}
for all $j\in \mathcal{K}\setminus \mathsf{W}$. Substituting~\eqref{eq:CSIlineII30} and~\eqref{eq:CSIlineII31} into~\eqref{eq:CSIlineII29}, 
\begin{equation}\label{eq:CSIlineII32}
\mathbb{P}(\brm{U}=U|\bsf{W}=\mathsf{W},\bsf{W}\in \bsf{S}) = \frac{1}{K}.	
\end{equation}

In the case (ii), we have
\begin{align}
& \mathbb{P}(\brm{U}=U|\bsf{W}=\mathsf{W},\bsf{W}\in \bsf{S}) \nonumber\\ 
& =\mathbb{P}(\brm{U}=U|\bsf{W}=\mathsf{W}, \bsf{S}=\{\mathsf{W},i\})\nonumber\\
&\quad\times \mathbb{P}(\bsf{S}=\{\mathsf{W},i\}|\bsf{W}=\mathsf{W},\bsf{W}\in \bsf{S})\nonumber\\
& = \frac{1}{K},	\label{eq:CSIlineII33}
\end{align} noting that by the model assumption,
\[\mathbb{P}(\bsf{S}=\{\mathsf{W},i\}|\bsf{W}=\mathsf{W},\bsf{W}\in \bsf{S})=\frac{1}{K-1},\] and by the design of the protocol, \[{\mathbb{P}(\brm{U}=U|\bsf{W}=\mathsf{W}, \bsf{S}=\{\mathsf{W},i\})}=\frac{K-1}{K}.\] 

From~\eqref{eq:CSIlineII32} and~\eqref{eq:CSIlineII33}, we can conclude that ${\mathbb{P}(\brm{U}=U|\bsf{W}=\mathsf{W},\bsf{W}\in \bsf{S})}$ does not depend on $\mathsf{W}$.\\ 

\subsubsection*{Case~2} Consider an arbitrary $U=\{U_1,U_2\}$. (Recall that $|U_1|=|U_2|=M-1$.) Take an arbitrary $\mathsf{W}\in \mathcal{K}$. There are two cases as follows: (i) $\mathsf{W}\in U_1\cup U_2$, and (ii) $\mathsf{W}\not\in U_1\cup U_2$. 

In the case (i), w.l.o.g., assume that $\mathsf{W}\in U_1$. Note that $\bsf{W}=\mathsf{W}$ and $\mathsf{W}\in U_1$ together imply that $\bsf{S} = \mathsf{W}\cup U_2$ (by the design of the protocol). Then, we have 
\begin{align}
& \mathbb{P}(\brm{U} =U|\bsf{W}=\mathsf{W},\bsf{W}\in \bsf{S}) \nonumber\\
& =\mathbb{P}(\brm{U}=U|\bsf{W}=\mathsf{W}, \bsf{S}=\mathsf{W}\cup U_2)\nonumber\\
&\quad \times \mathbb{P}(\bsf{S}=\mathsf{W}\cup U_2|\bsf{W}=\mathsf{W},\bsf{W}\in \bsf{S}).\label{eq:CSIlineII34}
\end{align} By the model assumption, we have 
\begin{equation}\label{eq:CSIlineII35}
\mathbb{P}(\bsf{S}=\mathsf{W}\cup U_2|\bsf{W}=\mathsf{W},\bsf{W}\in \bsf{S})=\binom{K-1}{M-1}^{-1}.	
\end{equation} Moreover, given that $\bsf{W}=\mathsf{W}$ and $\bsf{S} = \mathsf{W}\cup U_2$, the protocol constructs $U_1$ with probability $(\frac{2M-2}{K})\times\binom{K-M}{M-2}^{-1}$, noting that $\mathsf{W}\in U_1$. (The protocol selects the demand index $\mathsf{W}$ to be one of the elements in $U_1$ with probability $\frac{2M-2}{K}$, and selects the set of other $M-2$ elements in $U_1$ from the set of $K-M$ indices in $\mathcal{K}\setminus \mathsf{S}$ with probability $\binom{K-M}{M-2}^{-1}$.) This implies that
\begin{align}
& \mathbb{P}(\brm{U}=U|\bsf{W}=\mathsf{W},\bsf{S}=\mathsf{W}\cup U_2) \nonumber\\ 
& =2\left(\frac{M-1}{K}\right) \binom{K-M}{M-2}^{-1}.\label{eq:CSIlineII36}	
\end{align} Substituting~\eqref{eq:CSIlineII35} and~\eqref{eq:CSIlineII36} into~\eqref{eq:CSIlineII34}, 
\begin{align}
& \mathbb{P}(\brm{U}=U|\bsf{W}=\mathsf{W},\bsf{W}\in \bsf{S}) \nonumber\\
& = 2\left(\frac{M-1}{K}\right) \binom{K-M}{M-2}^{-1}\binom{K-1}{M-1}^{-1}.\label{eq:CSIlineII37}	
\end{align}

In the case (ii), we have
\begin{align}
& \mathbb{P}(\brm{U}=U|\bsf{W}=\mathsf{W},\bsf{W}\in \bsf{S}) \nonumber \\
& =\mathbb{P}(\brm{U}=U|\bsf{W}=\mathsf{W}, \bsf{S}=\mathsf{W}\cup U_1)\nonumber\\
& \quad\quad\times \mathbb{P}(\bsf{S}=\mathsf{W}\cup U_1|\bsf{W}=\mathsf{W},\bsf{W}\in \bsf{S})\nonumber\\
& \quad+\mathbb{P}(\brm{U}=U|\bsf{W}=\mathsf{W}, \bsf{S}=\mathsf{W}\cup U_2)\nonumber\\
& \quad\quad\times \mathbb{P}(\bsf{S}=\mathsf{W}\cup U_2|\bsf{W}=\mathsf{W},\bsf{W}\in \bsf{S})\nonumber\\
& = 2\left(1-\frac{2M-2}{K}\right)\binom{K-M}{M-1}^{-1}\binom{K-1}{M-1}^{-1},\label{eq:CSIlineII38}
\end{align} noting that 
\begin{align*}
&\mathbb{P}(\brm{U}=U|\bsf{W}=\mathsf{W},\bsf{S}=\mathsf{W}\cup U_1)\\ 
&=\mathbb{P}(\brm{U}=U|\bsf{W}=\mathsf{W},\bsf{S}=\mathsf{W}\cup U_2)\\ 
& =\left(1-\frac{2M-2}{K}\right)\binom{K-M}{M-1}^{-1},	
\end{align*} and 
\begin{align*}
& \mathbb{P}(\bsf{S}=\mathsf{W}\cup U_1|\bsf{W}=\mathsf{W},\bsf{W}\in \bsf{S})\\ 
&  =\mathbb{P}(\bsf{S}=\mathsf{W}\cup U_2|\bsf{W}=\mathsf{W},\bsf{W}\in \bsf{S})\\
&  = \binom{K-1}{M-1}^{-1}	.
\end{align*} Now, it is easy to verify that 
\begin{align*}
& \left(\frac{M-1}{K}\right)\binom{K-M}{M-2}^{-1} \\ 
& = \left(1-\frac{2M-2}{K}\right)\binom{K-M}{M-1}^{-1}.	
\end{align*}
This shows that~\eqref{eq:CSIlineII37} and~\eqref{eq:CSIlineII38} are equal, completing the proof that $\mathbb{P}(\brm{U}=U|\bsf{W}=\mathsf{W},\bsf{W}\in \bsf{S})$ does not depend on $\mathsf{W}$.\\ 
 
\subsubsection*{Case~3} Consider an arbitrary query $U=\{U_1,U_2\}$. (Recall that $|U_1|=|U_2|=M$.) Take an arbitrary $\mathsf{W}\in \mathcal{K}$. There are two cases as follows: (i) $\mathsf{W}\in U_1\cap U_2$, and (ii) $\mathsf{W}\not\in U_1\cap U_2$. 

In the case (i), we have
\begin{align}
& \mathbb{P}(\brm{U}=U|\bsf{W}=\mathsf{W},\bsf{W}\in \bsf{S})\nonumber \\ 
& =\mathbb{P}(\brm{U}=U|\bsf{W}=\mathsf{W}, \bsf{S}= U_1)\nonumber\\
&\quad\quad\times \mathbb{P}(\bsf{S}=U_1|\bsf{W}=\mathsf{W},\bsf{W}\in \bsf{S})\nonumber\\
&\quad+\mathbb{P}(\brm{U}=U|\bsf{W}=\mathsf{W}, \bsf{S}=U_2)\nonumber\\
&\quad\quad\times \mathbb{P}(\bsf{S}=U_2|\bsf{W}=\mathsf{W},\bsf{W}\in \bsf{S})\nonumber\\
& = 2\left(\frac{2M-K}{K}\right)\binom{M-1}{2M-K-1}^{-1}\binom{K-1}{M-1}^{-1}, \label{eq:CSIlineII39}
\end{align} noting that 
\begin{align*}
&\mathbb{P}(\brm{U}=U|\bsf{W}=\mathsf{W},\bsf{S}=U_1)\\ 
& =\mathbb{P}(\brm{U}=U|\bsf{W}=\mathsf{W},\bsf{S}=U_2)\\ 
& =\left(\frac{2M-K}{K}\right) \binom{M-1}{2M-K-1}^{-1},	
\end{align*} and
\begin{align*}
& \mathbb{P}(\bsf{S}=U_1|\bsf{W}=\mathsf{W},\bsf{W}\in \bsf{S})\\ 
& =\mathbb{P}(\bsf{S}=U_2|\bsf{W}=\mathsf{W},\bsf{W}\in \bsf{S})\\
& = \binom{K-1}{M-1}^{-1}.	
\end{align*} 

In the case (ii), w.l.o.g., assume that $\mathsf{W}\in U_1$. Then, we have
\begin{align}
& \mathbb{P}(\brm{U}=U|\bsf{W}=\mathsf{W},\bsf{W}\in \bsf{S}) \nonumber \\ 
&=\mathbb{P}(\brm{U}=U|\bsf{W}=\mathsf{W}, \bsf{S}=U_1)\nonumber\\
&\quad\times \mathbb{P}(\bsf{S}=U_1|\bsf{W}=\mathsf{W},\bsf{W}\in \bsf{S})\nonumber\\
& = 2\left(\frac{K-M}{K}\right) \binom{M-1}{2M-K}^{-1}\binom{K-1}{M-1}^{-1},\label{eq:CSIlineII40}
\end{align} noting that 
\begin{align*}
& \mathbb{P}(\brm{U}=U|\bsf{W}=\mathsf{W},\bsf{S}=U_1) \\ 
& =2\left(\frac{K-M}{K}\right) \binom{M-1}{2M-K}^{-1},	
\end{align*}
and 
\begin{align*}
\mathbb{P}(\bsf{S}=U_1|\bsf{W}=\mathsf{W},\bsf{W}\in \bsf{S}) = \binom{K-1}{M-1}^{-1}.	
\end{align*}
It is easy to verify that 
\begin{align*}
& \left(\frac{2M-K}{K}\right)\binom{M-1}{2M-K-1}^{-1} \\ 
& = \left(\frac{K-M}{K}\right)\binom{M-1}{2M-K}^{-1}.	
\end{align*} This shows that~\eqref{eq:CSIlineII39} and~\eqref{eq:CSIlineII40} are equal, completing the proof that $\mathbb{P}(\brm{U}=U|\bsf{W}=\mathsf{W},\bsf{W}\in \bsf{S})$ does not depend on $\mathsf{W}$.\\ 

\subsubsection*{Case~4} By the protocol, we have $U = \mathcal{K}$, and hence ${\mathbb{P}(\brm{U}=U|\bsf{W}=\mathsf{W},\bsf{W}\in \bsf{S})=1}$ for all $\mathsf{W}$.

\bibliographystyle{IEEEtran}
\bibliography{PIR_salim,pir_bib,coding1,coding2}

\end{document}